\documentclass{aa}
\usepackage{txfonts}
\usepackage{epsfig}
\usepackage{subfigure}
\usepackage{multirow}

\usepackage{times}
\usepackage{natbib}
\usepackage{longtable,lscape}
\bibpunct{(}{)}{;}{a}{}{,}

\begin{document}

\title{CAIXA: a Catalogue of AGN In the XMM-\textit{Newton} Archive\\ I. Spectral analysis}
\author{Stefano Bianchi\inst{1,2}, Matteo Guainazzi\inst{2}, Giorgio Matt\inst{1}, Nuria Fonseca Bonilla\inst{2}, Gabriele Ponti\inst{3}}

\offprints{Stefano Bianchi\\ \email{bianchi@fis.uniroma3.it}}

\institute{Dipartimento di Fisica, Universit\`a degli Studi Roma Tre, via della Vasca Navale 84, 00146 Roma, Italy
\and XMM-Newton Science Operations Center, European Space Astronomy Center, ESA, Apartado 50727, E-28080 Madrid, Spain
\and Laboratoire APC, UMR 7164, 10 rue A. Domon et L. Duquet, 75205 Paris, France}

\date{Received / Accepted}

\authorrunning{S. Bianchi et al.}

\abstract
{}
{We present CAIXA, a Catalogue of AGN In the XMM-\textit{Newton} Archive. It consists of all the radio-quiet X-ray unobscured ($\mathrm{N_H}<2\times10^{22}$ cm$^{-2}$) Active Galactic Nuclei (AGN) observed by XMM-\textit{Newton} in targeted observations, whose data are public as of March 2007. With its 156 sources, this is the largest catalogue of high signal-to-noise X-ray spectra of AGN.}
{All the EPIC pn spectra of the sources in CAIXA were extracted homogeneously and a baseline model was applied in order to derive their basic X-ray properties. These data are complemented by multiwavelength data found in the literature: Black Hole masses, Full Width Half Maximum (FWHM) of H$\beta$, radio and optical fluxes.}
{Here we describe our homogeneous spectral analysis of the X-ray data in CAIXA and present all the results on the parameters adopted in our best-fit models.}
{}


\maketitle

\section{Introduction}

Despite the unquestionable progress made in our understanding of the physics of Active Galactic Nuclei (AGN), there are still many open issues related to the X-ray properties of these objects. Some of the most discussed ones in the last years are: the highly debated nature of the soft X-ray excess in unobscured sources; the different spectral and timing properties for different classes of sources; the nature of radio emission in radio-quiet objects and its relation to the X-ray nuclear emission; the origin of the reprocessing of the primary emission from highly ionized material; the correlation of the above-mentioned phenomenology with fundamental properties of AGN, such as the Black Hole (BH) mass and the accretion rate. The most effective way to address these questions is to analyze large numbers of AGN with good-quality X-ray spectra and to perform statistical studies, taking into account data in other wavelengths and other basic properties of the objects.

The European Space Agency's (ESA) X-ray Multi-Mirror Mission (XMM-\textit{Newton}) was launched on December 10th 1999. The XMM-\textit{Newton} public archive has since become the repository of an enormous amount of high-quality X-ray data and a precious legacy for future missions. In particular, our knowledge of the physics of AGN has dramatically improved thanks to the large sensitivity of the European Photon Imaging Camera (EPIC) pn charge-coupled device (CCD) arrays. At this time, a systematic and homogeneous study of the EPIC pn spectra of AGN represents a necessary step to fully take advantage of this highly successful X-ray mission.

In this paper, we present CAIXA, a Catalogue of AGN In the XMM-\textit{Newton} Archive. This catalogue was already used by \citet{bianchi07} to confirm with much higher confidence the `Iwasawa-Taniguchi' effect (i.e. the anti-correlation between the Equivalent Width of the neutral iron narrow emission line and the X-ray luminosity) and by \citet[][a smaller catalogue was used at that period]{gbd06} and \citet{long07}, to assess the frequency of the relativistic component of the iron line. Here we describe our homogeneous spectral analysis of the X-ray data in CAIXA and present all the results on the parameters adopted in our best-fit models. In following papers, we will present the timing analysis and properties of CAIXA and we will investigate the correlations between the X-ray and the multiwavelength properties of the sources in the catalogue.

\section{The catalogue}

\subsection{\label{data}X-ray data reduction and spectral analysis}

CAIXA consists of all the radio-quiet X-ray unobscured ($\mathrm{N_H}<2\times10^{22}$ cm$^{-2}$) AGN observed by XMM-\textit{Newton} in targeted observations, whose data are public as of March 2007. Only EPIC pn \citep{struder01} data were reprocessed, with SASv6.5 \citep{sas610}. For the observations performed in Small Window mode, background spectra were generated using blank-field event lists, according to the procedure presented in \citet{rp03}. In all other cases, background spectra were extracted from source-free regions close to the target in the observation event file.

Source extraction radii and screening for intervals of flaring particle background were performed via an iterative process which leads to a maximization of the signal-to-noise ratio, similarly to what described in \citet{pico04}. Spectra were binned in order to oversample the intrinsic instrumental energy resolution by a factor not lower than 3, and to have spectral bins with at least 25 background-subtracted counts. This ensures the applicability of the $\chi^2$ statistics.

We applied a number of criteria to filter the catalogue. First of all, spectra with less than 200 counts in either of the (rest-frame) bands of 0.5-2 and 2-10 keV were rejected, because they do not possess enough independent bins to be fitted with our models. Moreover, spectra affected by a pileup larger than 1$\%$ were rejected. Only two objects, namely NGC~2992 and ESO~548-G081, did not have an alternative pileup-free observation and were therefore excluded from the catalogue. After the fitting procedure, all sources with a local column density (measured in the 2-10 keV band) larger than $2\times10^{22}$ cm$^{-2}$ were further excluded from the catalogue.

In order to ensure that only radio-quiet object were included in CAIXA, we collected radio data at 6 cm (4.85 GHz) and 20 cm (1.4 GHz) from the literature and calculated the \textit{K}-corrected radio-loudness parameter \textit{R} \citep[$\log R =\log f_\mathrm{6\,cm} - \log f_B$:][]{sto92} and the X-ray radio-loudness parameter $R_X$ \citep[$\log R_X =\log f_\mathrm{6\,cm} - \log f_\mathrm{2-10\,keV}$:][]{tw03}. When only a 20 cm radio flux was available, it was converted to 6 cm, assuming a power law radio spectrum ($f_\nu\propto\nu^{-\alpha_\mathrm{r}}$), where $\alpha_\mathrm{r}=0.7$ is the average two-point radio spectral index for the sources in CAIXA with measures both at 6 and 20 cm (61 sources). The same procedure was followed for the (few) 843 MHz measures. The optical flux $f_B$ was derived from the \textit{V} magnitude and the colour index \textit{B-V}, the latter assumed to be 0.3 if not available in literature. According to the standard definition, all quasars with $\log(R)>1$ were excluded, while for Seyfert galaxies, known to be on average `radio-louder', $\log(R)>2.4$ and $\log(R_X)>-2.755$ were used as boundaries  \citep[see][]{panessa07}. Note that three quasars are nominally radio-loud, having values of $R$ of 1.09 (RXJ0136.9-3510), 1.01 (RXJ0228.2-4057) and 1.08 (KUV18217+6419). However, given that the uncertainties on the radio and optical fluxes are not taken into consideration, we are confident these sources can be safely considered radio-quiet. Moreover, we did not find radio measures for the following sources:  ESO~198-G24, RXJ0323.2-4931, Fairall~1116, 1H0707-495, J124126.3-574955, H1846-786, RXJ2241.8-4405, REJ2248-511. Some of them are treated as radio-quiet sources in papers cited in Table \ref{multitable} and, in any case, the vast majority of them are Seyfert galaxies and unlikely to be radio-loud with our criteria. Therefore, we kept all of them in our final catalogue. Finally, with respect to the catalogue used in \citet{bianchi07}, we excluded PMNJ0623-6436, for which we found a new radio observation at 6 cm \citep{sad06}, leading to $\log R\simeq2$. Nevertheless, this exclusion does not affect the results published in \citet{bianchi07}.

At the end of this selection procedure, the total catalogue comprises 156 radio-quiet AGN. When more than one observation of the same object was available, we used the one with the longest exposure, independently of its flux status, in order to avoid, as much as possible, any bias. Tables \ref{xraydata} and \ref{xraydata2} summarize the properties of the XMM-\textit{Newton} observations included in CAIXA, along with the fluxes, luminosities and main X-ray parameters derived from the fits with the baseline model.

The baseline model consists of the following components:

\begin{equation}
F\left(E\right) = e^{-\sigma\left(E\right)N_H^G}\left[Ae^{-\sigma\left(E\right)N_H^s}E^{-\Gamma_s} + Be^{-\sigma\left(E\right)N_H^h}E^{-\Gamma_h} + 4\times G\right]
\end{equation} 

\noindent where $\sigma\left(E\right)$ is the photoelectric cross-section, adopting solar abundances as in \citet{ae82}, $N_H^G$ is the Galactic column density appropriate for each source \citep[after][]{dl90}, $N_H^s<N_H^h$ are two local column densities at the redshift of the source, possibly coexisting if related to absorption on different scales, $A$ and $B$ are two normalization factors, $\Gamma_s$ and $\Gamma_h$ the spectral photon indexes for the soft an the hard spectrum, and $G$ are four Gaussian emission lines at rest-frame energies fixed to 6.4, 6.7, 6.96 and 7.06 keV, as appropriate for neutral, He- and H-like iron K$\alpha$ and neutral iron K$\beta$, respectively. The inclusion of the latter allows us to disentangle its possible contamination to the 6.96 keV line, by not allowing its flux to be larger than 0.16 times the flux of the neutral K$\alpha$ line \citep{mbm03}. A relativistically broadened component of the iron line was never included in our fits \citep[see][for a discussion on this point]{bianchi07}. Where the addition of the second powerlaw leads to a $\chi^2$ improvement less than 99 \% (according to F-test), we put $B=0$ and the baseline model only includes a powerlaw component. Finally, two absorption edges at 0.737 (O \textsc{vii}) and 0.871 (O \textsc{viii}) keV are included when required, to roughly model a warm absorber (see also Sect. \ref{goodfit} and \ref{gamma} for a discussion on this issue). No Compton reflection component is included in our fits, since the limited bandpass of XMM-\textit{Newton} does not allow us to give good constraints on this parameter, in particular when the SNR is not very large. We will discuss the possible consequences of this choice on our results in Sect. \ref{gamma}. In order to study the correlations where the soft X-ray luminosities are involved, we decided to compile a sub-catalogue, where all the sources with a column density in excess of the Galactic one (but still lower than the general $2\times10^{22}$ cm$^{-2}$ limit of CAIXA) were excluded. This sub-catalogue comprises 142 sources and will be explicitly referred to when used in this paper instead of the whole CAIXA.

All spectra were analyzed with \textsc{Xspec} 12.2.1 \citep{xspec}. In the following, errors correspond to the 90\% confidence level for one interesting parameter ($\Delta \chi^2 =2.71$), where not otherwise stated. The cosmological parameters used throughout this paper are $H_0=70$ km s$^{-1}$ Mpc$^{-1}$, $\Omega_\Lambda=0.73$ and $\Omega_m=0.23$ (i.e. the default ones in \textsc {Xspec} 12.2.1).

\addtocounter{table}{1}

\subsection{\label{multi}Multiwavelength data and BH masses}

Table \ref{multitable} summarizes the multiwavelength data collected for the sources in CAIXA, i.e. the absolute magnitude (100\% coverage), the radio fluxes at 6 and/or 20 cm (95\% coverage), Full Width Half Maximum (FWHM) of H$\beta$ (65\% coverage), BH mass (57\%), \textit{V} magnitude (100\% coverage) and \textit{B-V} colour index (65\% coverage). All the objects of the catalogue are included in \citet{ver06}, who represented the main reference for the multiwavelength data. All the other references are shown in Table \ref{multitable}. We stress here that the FWHM of H$\beta$ in CAIXA are all larger than 650 km s$^{-1}$, as expected since it includes only sources unobscured or moderately obscured in the X-rays. Therefore, our distinction between narrow- and broad-line objects simply reflects the standard limit of H$\beta$ FWHM$<2\,000$ km s$^{-1}$, while no optical Type 2 objects are present in CAIXA. Moreover, the standard boundary of M$_\mathrm{abs}=-23$ was adopted to classify a source as a quasar or as a Seyfert galaxy.

\addtocounter{table}{2}

\subsection{General properties}

\begin{figure*}
\begin{center}
\epsfig{file=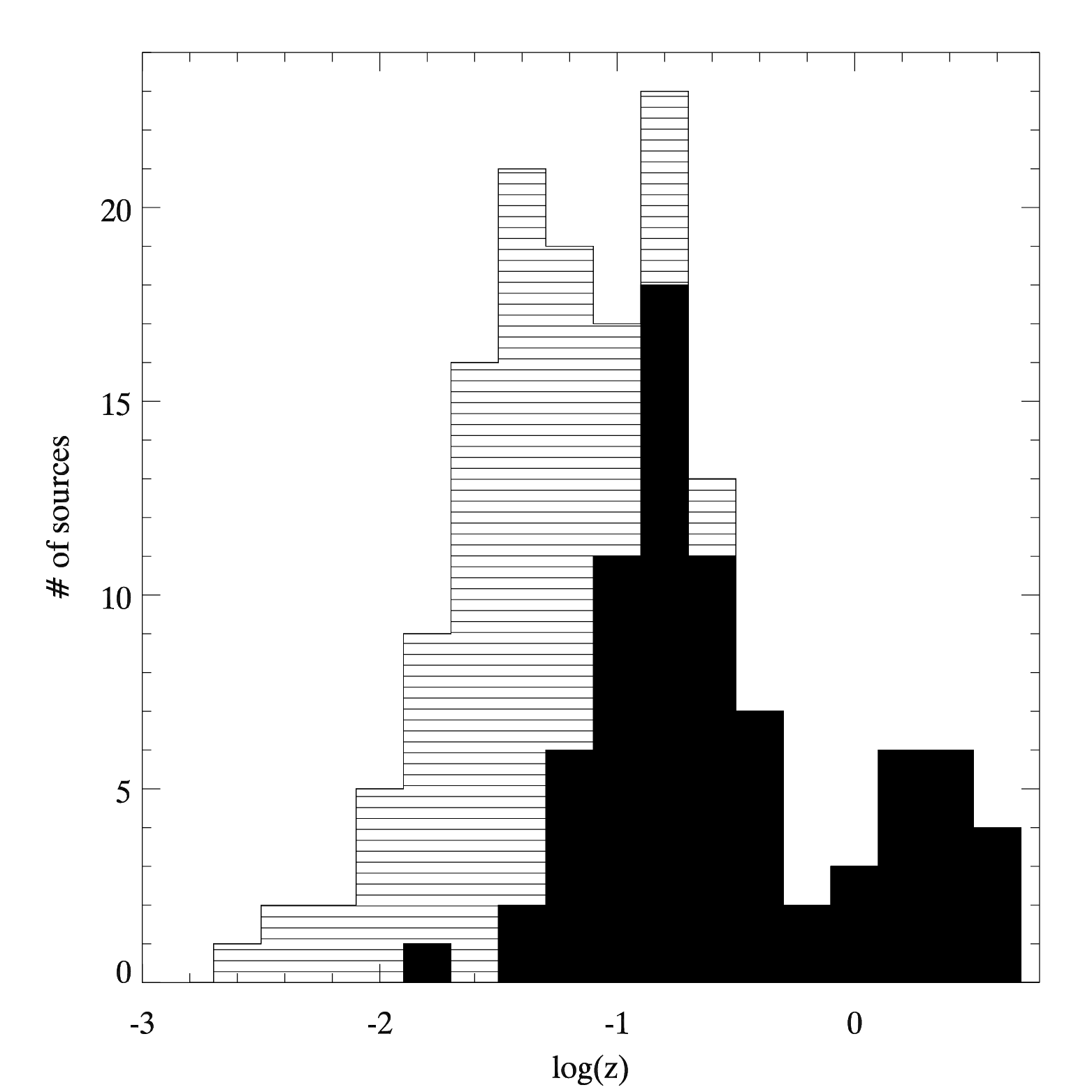, width=8cm, height=8cm}
\epsfig{file=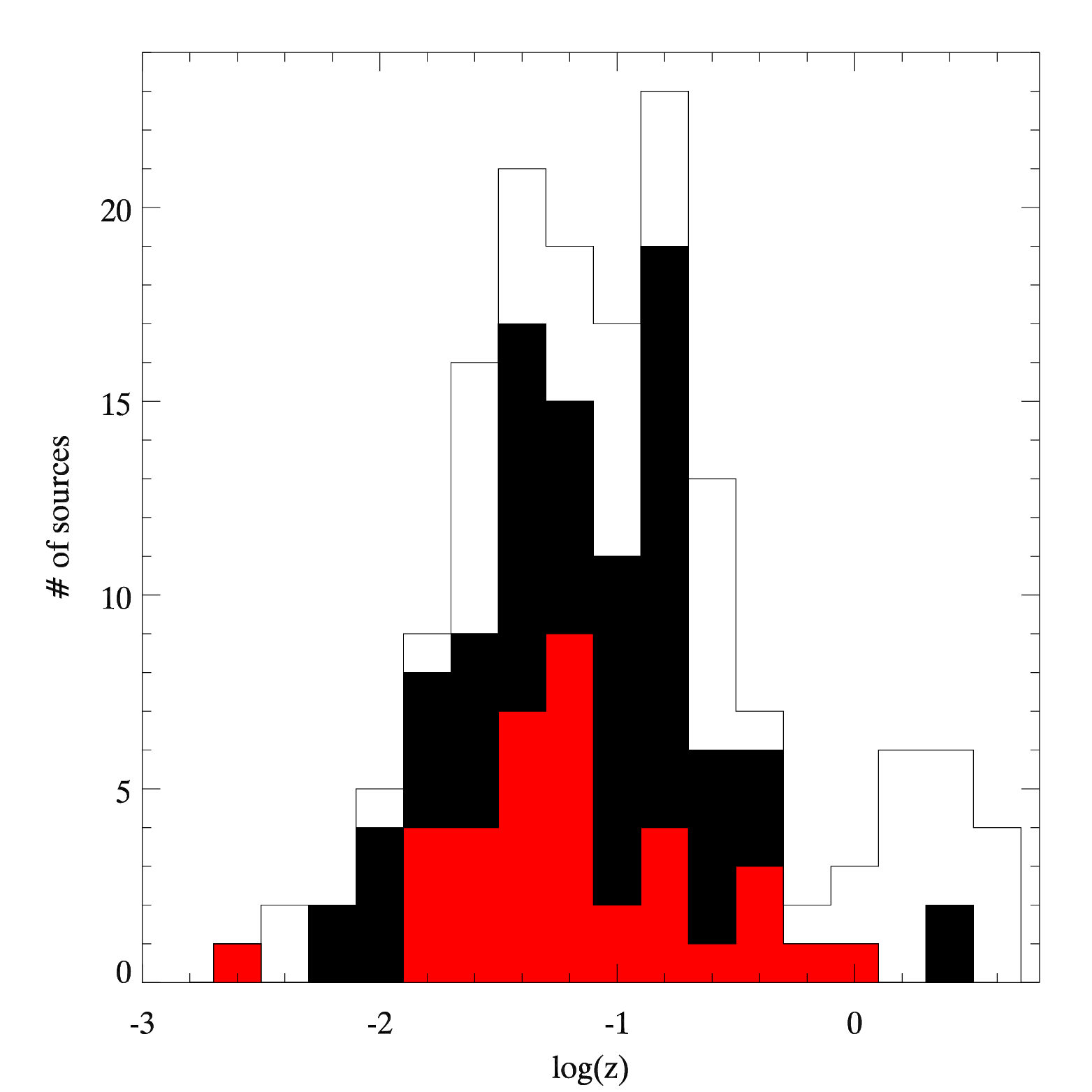, width=8cm, height=8cm}
\epsfig{file=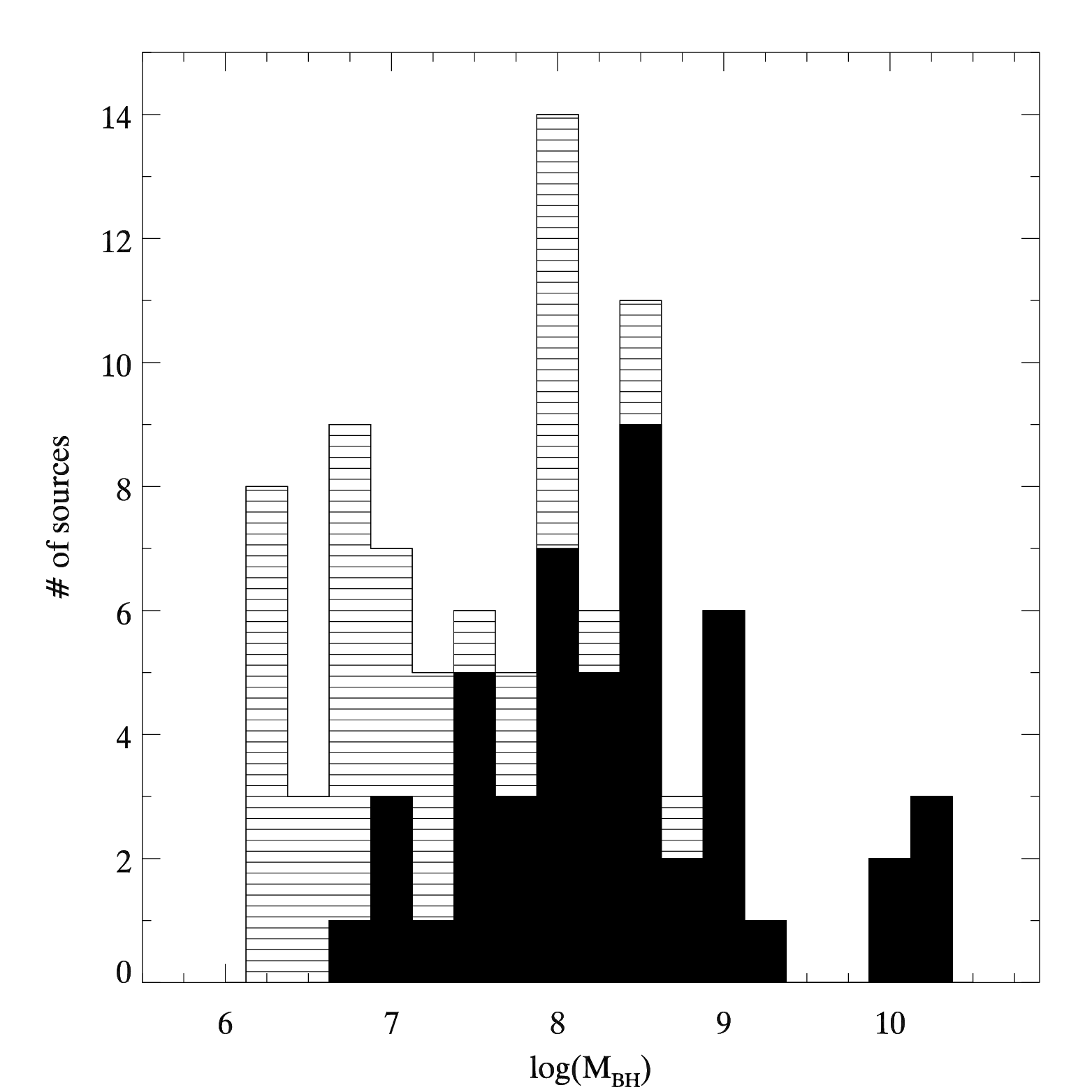, width=8cm, height=8cm}
\epsfig{file=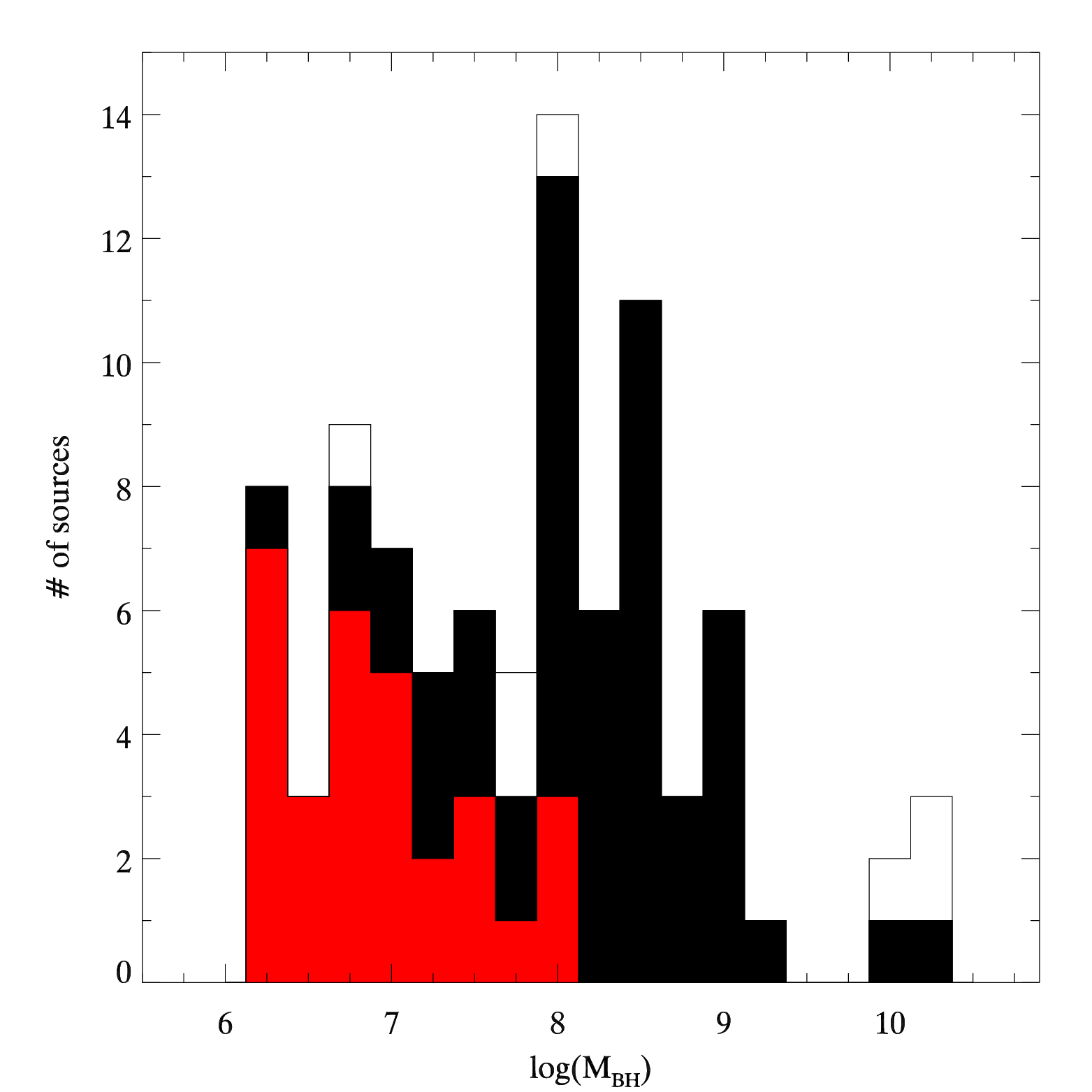, width=8cm, height=8cm}
\end{center}
\caption{\label{zdist}General properties of the sources included in CAIXA. Redshift (\textit{top}) and BH mass (\textit{bottom}) distributions. Quasars (filled) and Seyferts (dashed) are shown on the left, while narrow-line objects (red), broad-line objects (black) and objects without a measure of the H$\beta$ FWHM (white) are shown on the right. See text for details.}
\end{figure*}

CAIXA includes a total of 77 quasars (16 narrow-line and 35 broad-line objects) and 79 Seyfert galaxies (21 narrow-line and 30 broad-line objects). The redshift distribution, which spans from z=0.002 to z=4.520, is shown in Fig. \ref{zdist}: given the requirement of a high SNR X-ray spectrum, most of the sources are local (almost 90\% within $\mathrm{z}<1$). Also shown in the same figure are the BH masses distribution ($1.4\times10^6\,\mathrm{M_\odot}<\mathrm{M_{BH}}<2.0\times10^{10}\,\mathrm{M_\odot}$): as expected, quasars have, on average, larger BH masses and narrow-line objects the smallest ones, given the correlation between the BH mass and the FWHM of the H$\beta$.

\begin{figure*}
\begin{center}
\epsfig{file=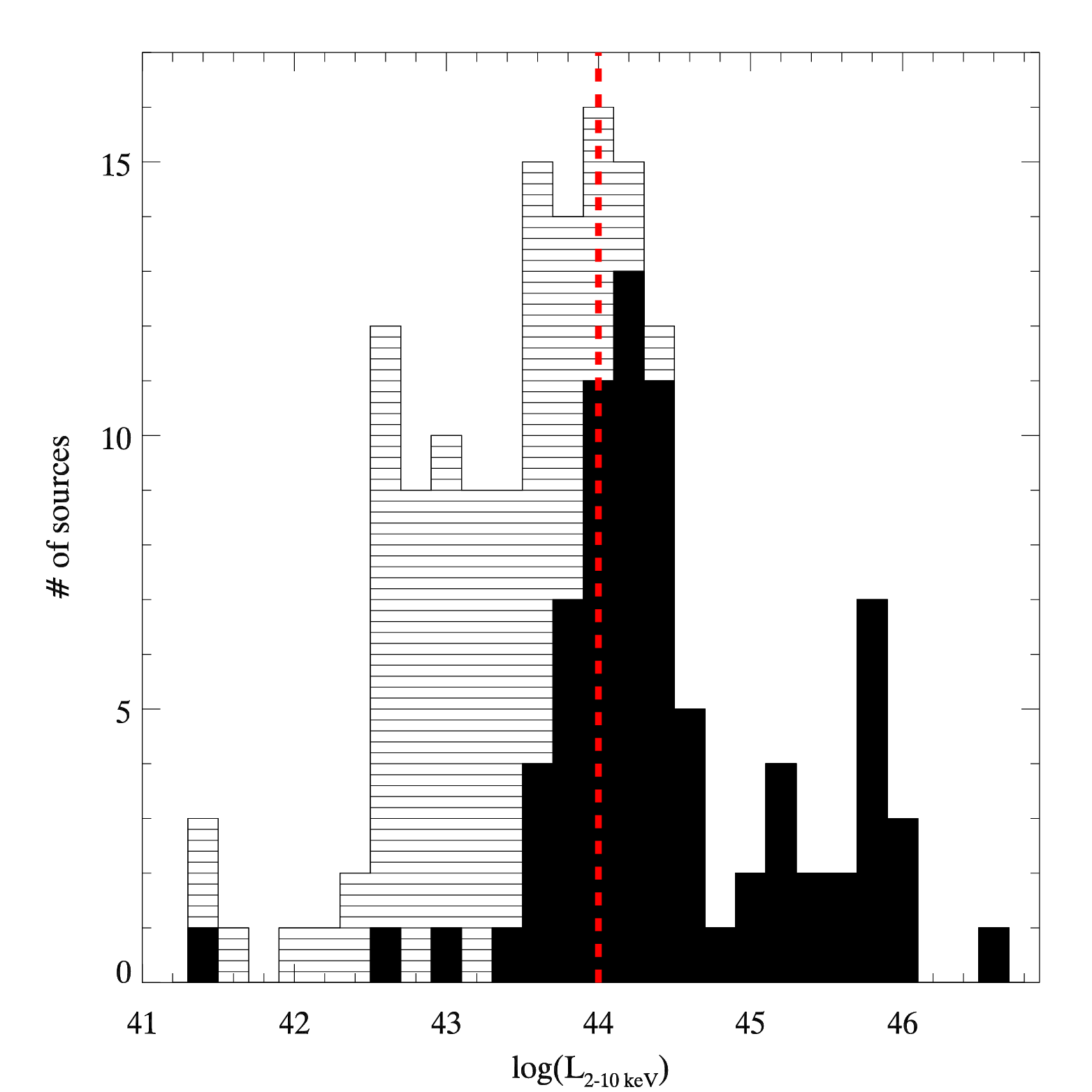, width=8cm, height=8cm}
\epsfig{file=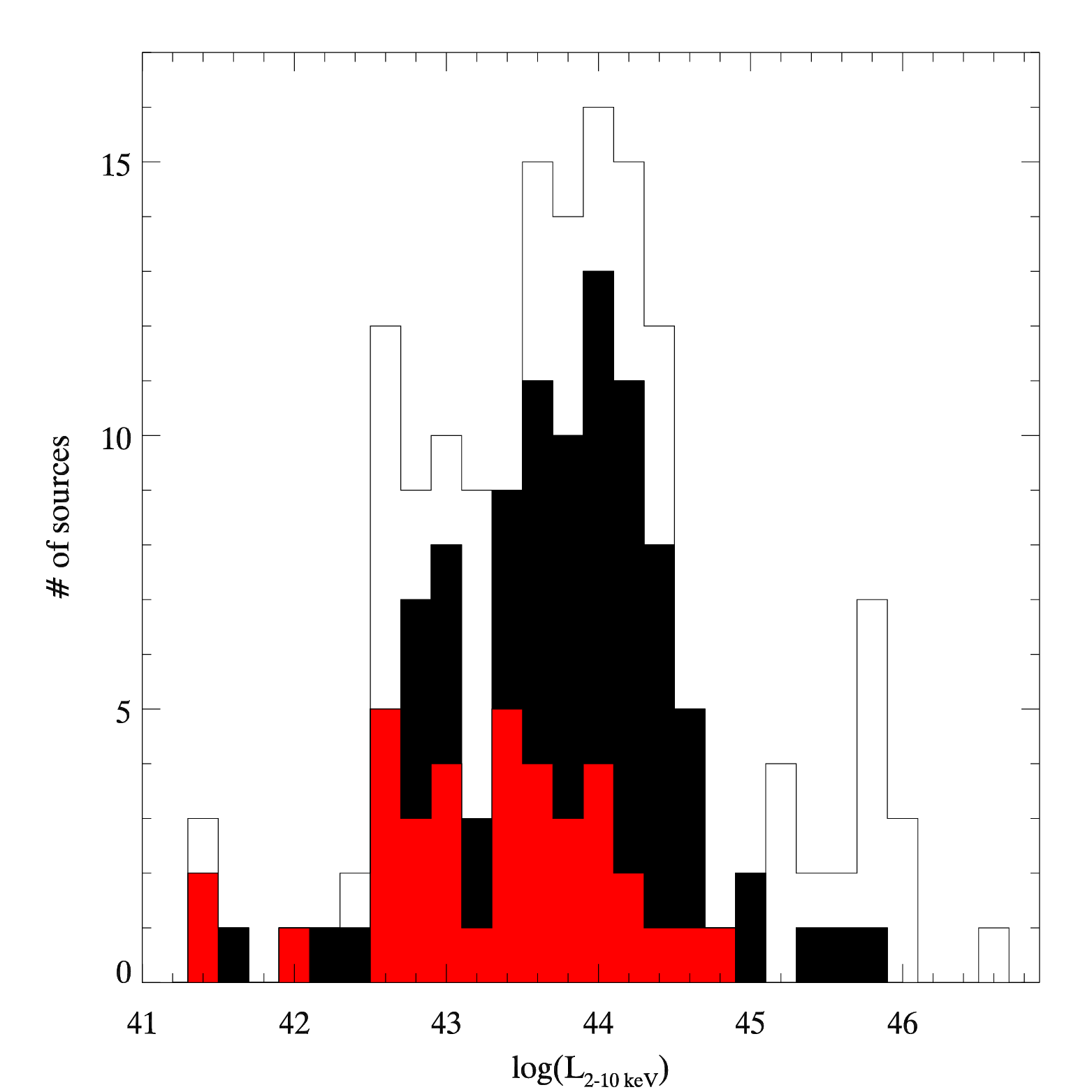, width=8cm, height=8cm}
\epsfig{file=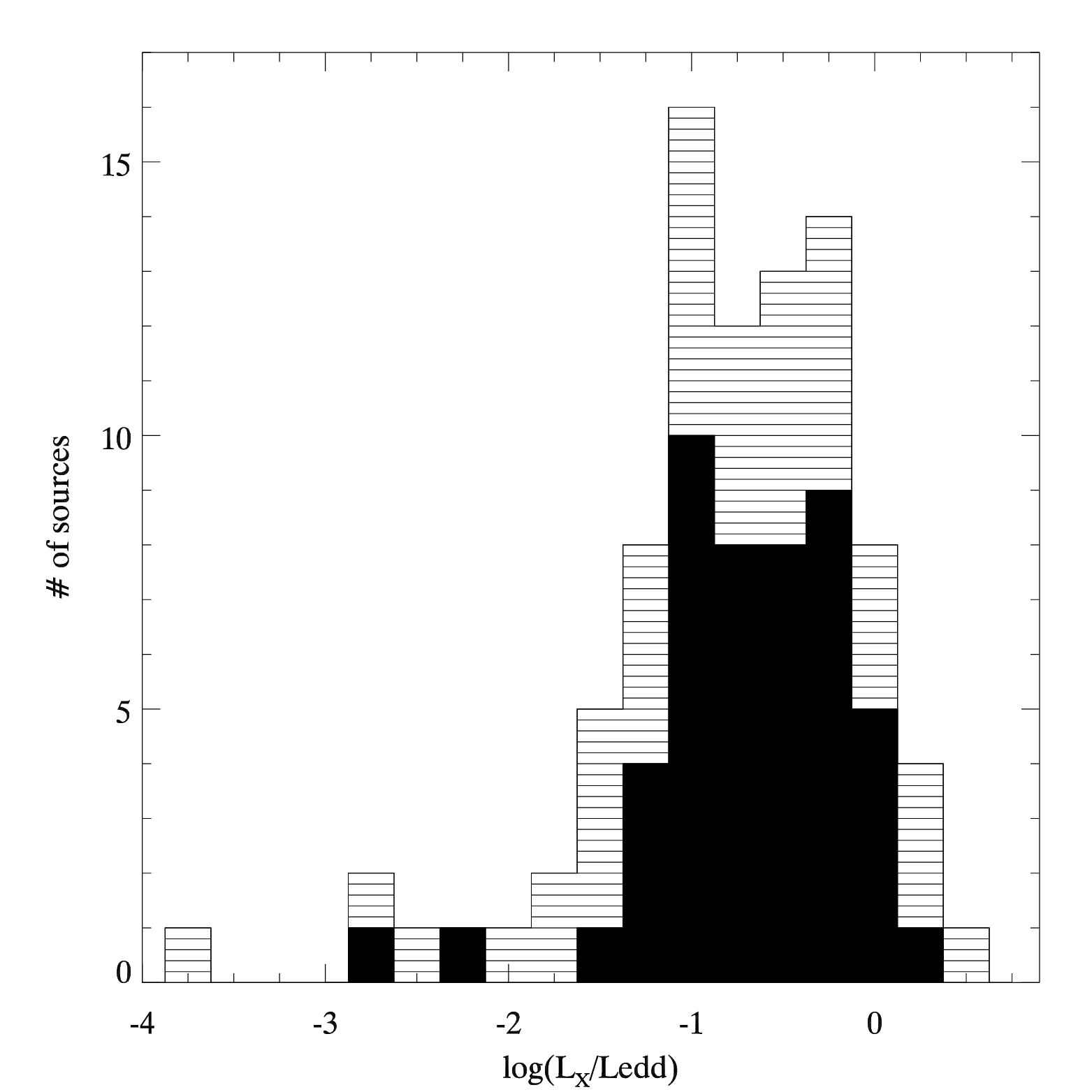, width=8cm, height=8cm}
\epsfig{file=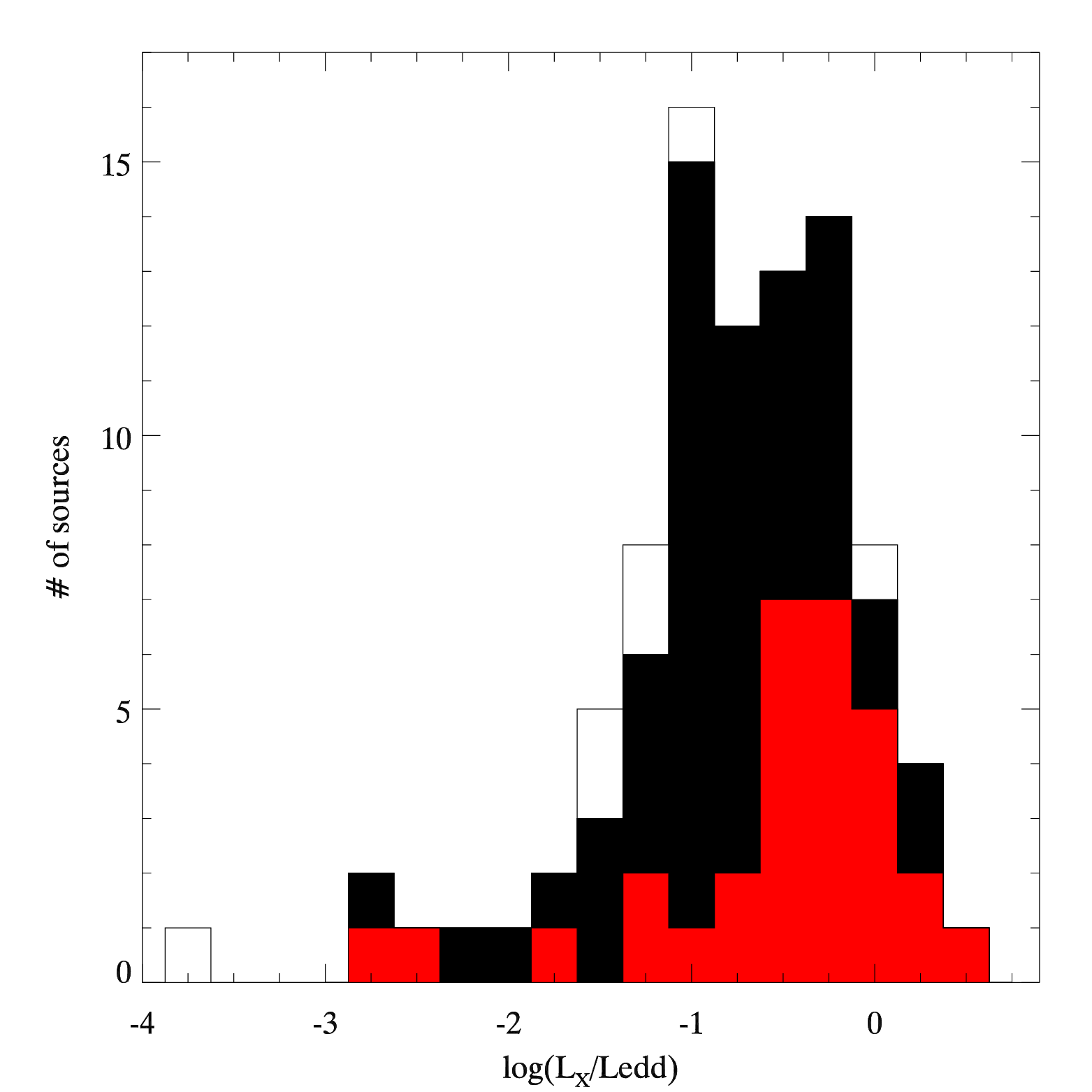, width=8cm, height=8cm}
\end{center}
\caption{\label{lxdist}General properties of the sources included in CAIXA. Hard X-ray (2-10 keV) luminosity (\textit{top}) and Eddington ratio (\textit{bottom}) distributions. Quasars (filled) and Seyferts (dashed) are shown on the left, while narrow-line objects (red), broad-line objects (black) and objects without a measure of the H$\beta$ FWHM (white) are shown on the right. See text for details.}
\end{figure*}

The distribution of the hard X-ray luminosities, which cover a range between $L_{2-10\,\mathrm{keV}}=2.0\times10^{41}$ to $3.9\times10^{46}$ erg s$^{-1}$, is shown in Fig. \ref{lxdist}. The customary X-ray boundary between quasars and Seyferts at $L_{2-10\,\mathrm{keV}}=1\times10^{44}$ erg s$^{-1}$ seems to be in fair agreement with the historical optical classification adopted in CAIXA, even if there is a larger number of X-ray weak quasars. Finally, the distribution of the Eddington ratios are shown in the lower panels of Fig. \ref{lxdist}: these values are derived from the 2-10 keV luminosities, adopting a luminosity-dependent bolometric correction, after \citet{mar04}. The average Eddington ratio is $\log(L_\mathrm{bol}/L_\mathrm{Edd})=-0.79\pm0.08$ (standard deviation $\sigma=0.71$) for all the catalogue, $-0.69\pm0.08$ ($\sigma=0.55$) for quasars, $-0.90\pm0.13$ ($\sigma=0.86$) for Seyferts, $-0.56\pm0.13$ ($\sigma=0.73$) for narrow-line objects and $-0.84\pm0.08$ ($\sigma=0.58$) for broad-line objects. These values are in agreement with other samples and catalogues \citep[e.g.][]{md04,bz04,nt07}, even if in CAIXA narrow-line objects do not show, on average, very large values.

\section{\label{results}Results}

\subsection{\label{goodfit}Goodness of fits}

\begin{figure}[!ht]
\begin{center}
\epsfig{file=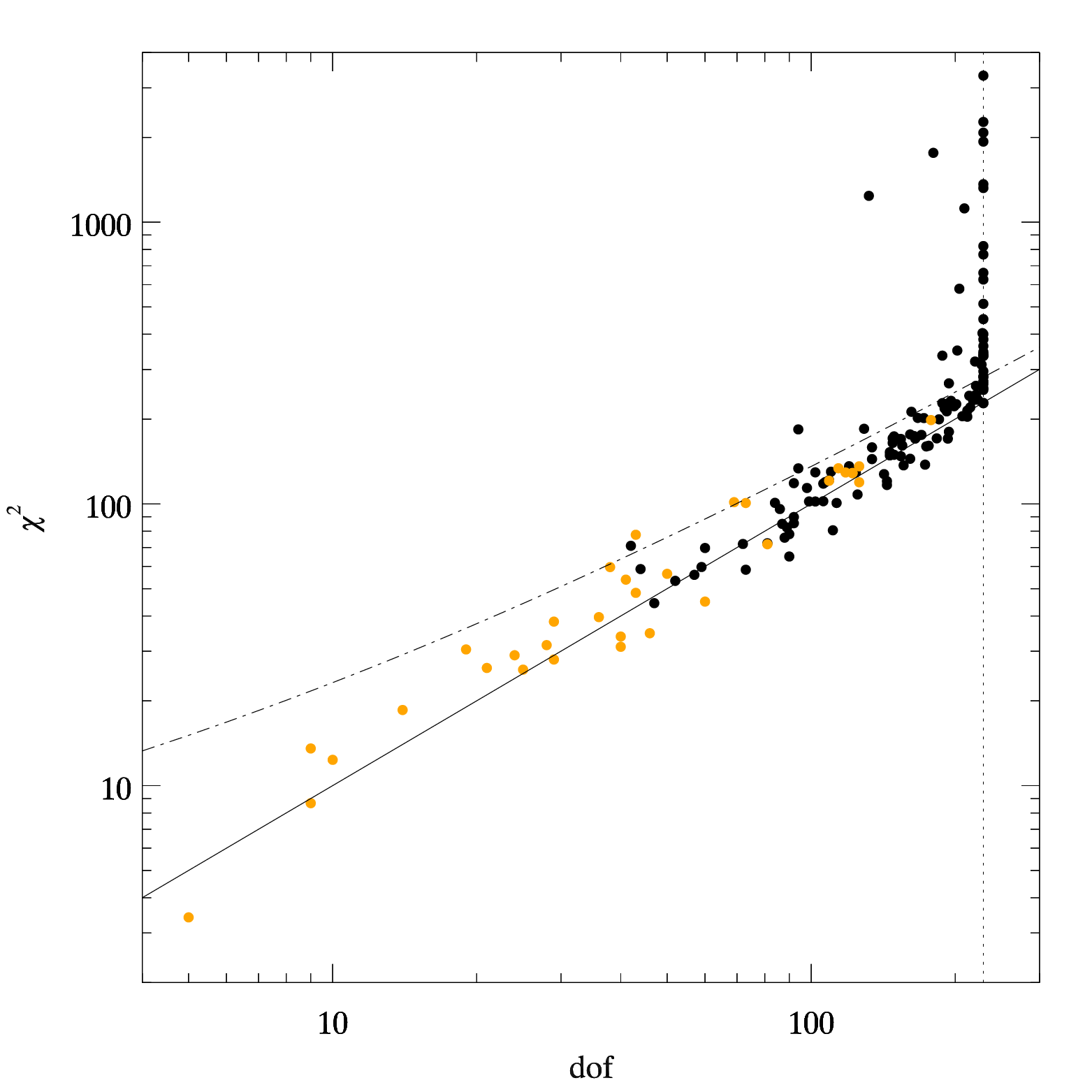, width=9cm, height=8cm}
\end{center}
\caption{\label{chidof}$\chi^2$ values obtained from the automatic fit procedure of CAIXA as a function of the degrees of freedom (dof) of each observation. Orange points are those where the second powerlaw is not statistically required. The continuous line refers to $\chi^2$/dof=1, while the dot-dashed curve the $\chi^2$ value corresponding to a 99\% Null Hypothesis Probability for the corresponding dof value. The vertical dotted line represents the 229 dof limit derived from the spectral binning criteria in CAIXA. See text for details.}
\end{figure}

\begin{figure*}
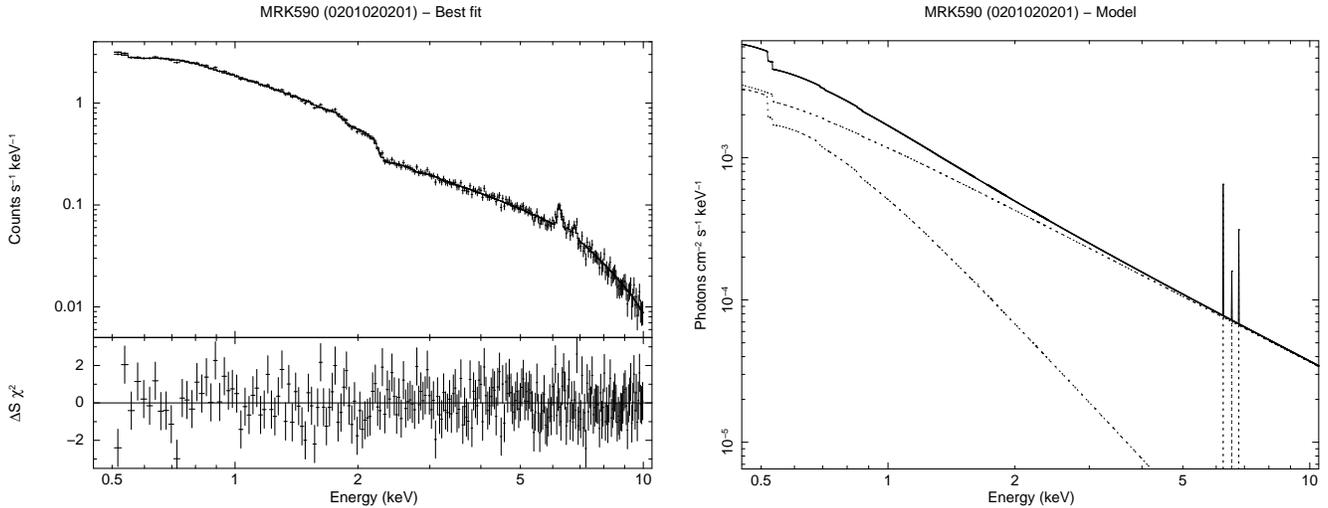

\begin{center}
\epsfig{file=mrk590_delchi.ps, width=6.8cm, angle=-90}
\hspace{0.2cm}
\epsfig{file=mrk590_model.ps, width=6.8cm, angle=-90}
\end{center}
\caption{\label{bestfit}Best fit data and model for a good fit in CAIXA, Mrk~590. Spectral parameters derived from this automatic fit are statistically reliable.}
\end{figure*}

Fig. \ref{chidof} shows the $\chi^2$ values obtained from the automatic fit procedure of CAIXA as a function of the degrees of freedom (dof) of each observation. The presence of a second powerlaw to model the soft X-ray emission is statistically required in 123 spectra out of 156 (see also Sect. \ref{data}): the single-powerlaw fits are marked in orange in Fig. \ref{chidof}. As expected, they are clustered in the low-dof part of the plot. The region where the adopted model can be considered correct at the 99\% confidence level is also shown. In other words, for each dof value there is a 1\% probability of getting a value of $\chi^2$ as large or larger than the dot-dashed curve plotted in figure if the model is correct. Most of the fits in CAIXA lie under this curve, confirming the good quality of the derived spectral parameter. As an example, we show in Fig. \ref{bestfit} the best fit for Mrk~590 ($\chi^2$=228/229), which, with its 229 dof (the maximum value allowed by the spectral binning criteria described in Sect. \ref{data}) is one of the observations with the highest SNR.

\begin{figure*}
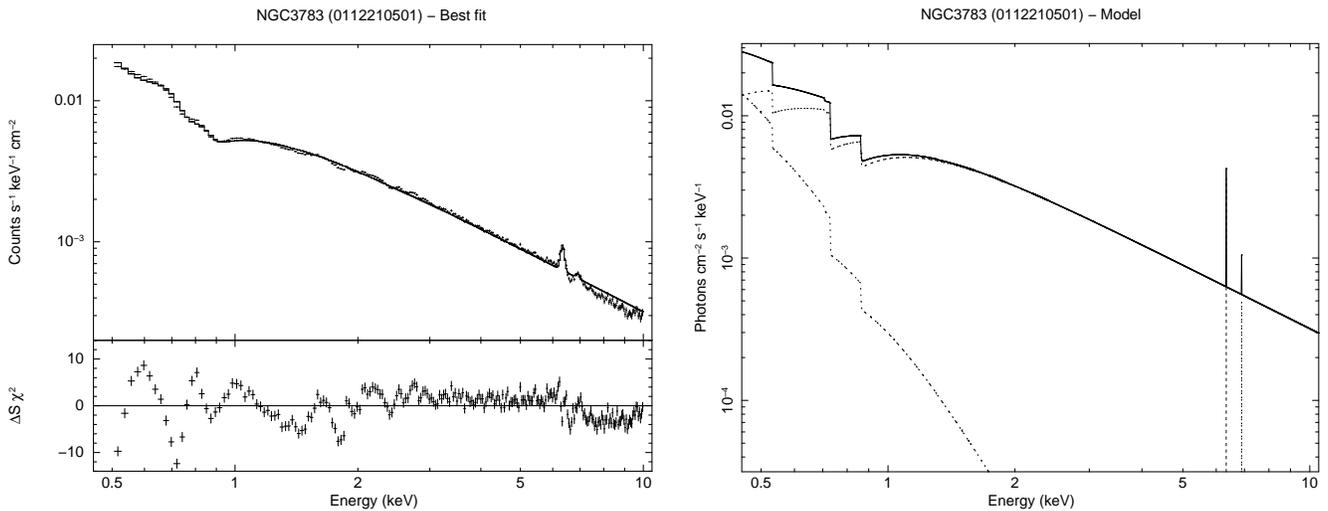

\begin{center}
\epsfig{file=ngc3783_delchi.ps, width=6.8cm, angle=-90}
\hspace{0.2cm}
\epsfig{file=ngc3783_model.ps, width=6.8cm, angle=-90}
\end{center}
\caption{\label{worstfit}Best fit data and model for a bad fit in CAIXA, NGC~3783. The spectral parameters derived from this automatic fit are still consistent with those extracted from a proper modeling of the data. See text for details.}
\end{figure*}

However, there is a number of fits whose $\chi^2$ cannot be considered statistically acceptable. Not surprisingly, most of them correspond to very large dof values: high SNR spectra are more difficult to be properly fitted by the simple model we are using in CAIXA. As an example, we show in Fig. \ref{worstfit} the ``best fit'' for NGC~3783 ($\chi^2$=2268/229), which, together with MCG-6-30-15, is one with the poorest reduced $\chi^2$ in CAIXA. As apparent from the left panel of Fig. \ref{worstfit}, the main problem is represented by huge residuals in the soft X-ray part of the spectrum, where the effects of absorption from ionised matter is only roughly taken into account by two absorption edges (see Sect. \ref{data}).

The lack of a proper modeling of warm absorbers in CAIXA is dictated by the complexity of these components, which do not allow an automatic handling. But, since we are not extracting information on this component in our catalogue, we should only check how much the spectral properties we are interested in are affected by this issue. In order to verify that, we fitted the NGC~3783 data with the model adopted by \citet{reev04}, which includes three phases for the warm absorbers, a thermal black body for the soft excess, a powerlaw and a Compton reflection component for the hard X-rays and the iron lines. The resulting $\chi^2$ is acceptable (303/221) and comparable with the one found by \citet{reev04}. The 0.5-2 keV flux is $16.06\pm0.02\times10^{-12}$ erg cm$^{-2}$ s$^{-1}$, while the 2-10 keV flux is $55.29\pm0.06\times10^{-12}$ erg cm$^{-2}$ s$^{-1}$. These values should be compared to the ones automatically calculated in CAIXA: $15.34\pm0.12\times10^{-12}$ erg cm$^{-2}$ s$^{-1}$, while the 2-10 keV flux is $55.48\pm0.15\times10^{-12}$ erg cm$^{-2}$ s$^{-1}$. Therefore, the difference in the soft X-ray flux is lower than 5\%, while the hard X-ray fluxes are consistent, within errors, and their nominal difference is of the order of 3\%. The EWs of the three iron lines are all consistent with the results obtained in CAIXA: $91^{+5}_{-4}$ ($93^{+4}_{-5}$) eV, $<1$ ($<1$) eV and $25\pm5$ ($17^{+3}_{-6}$) eV, respectively for neutral iron, Fe \textsc{xxv} and Fe \textsc{xxvi}.

We therefore conclude that, even when the statistical significance of the best fit in CAIXA is very low, as in the case presented above, the values of the X-ray parameters derived from the fit are still very reliable and can be safely used to properly characterise the sources. The only parameter which can be significantly affected by the simplicity of the model adopted in CAIXA is the photon spectral index. In the case of NGC~3783, for example, the value derived from the automatic fit procedure is very flat, being 1.49. The addition of a Compton reflection component (with R=1, consistent with the neutral iron K$\alpha$ line, as suggested by \citet{reev04}) and a complex warm absorber significantly steepens the powerlaw index to a more canonical value, 1.76. We will discuss this issue in the next section.

\subsection{\label{gamma}The X-ray spectral index}

\begin{figure*}
\begin{center}
\epsfig{file=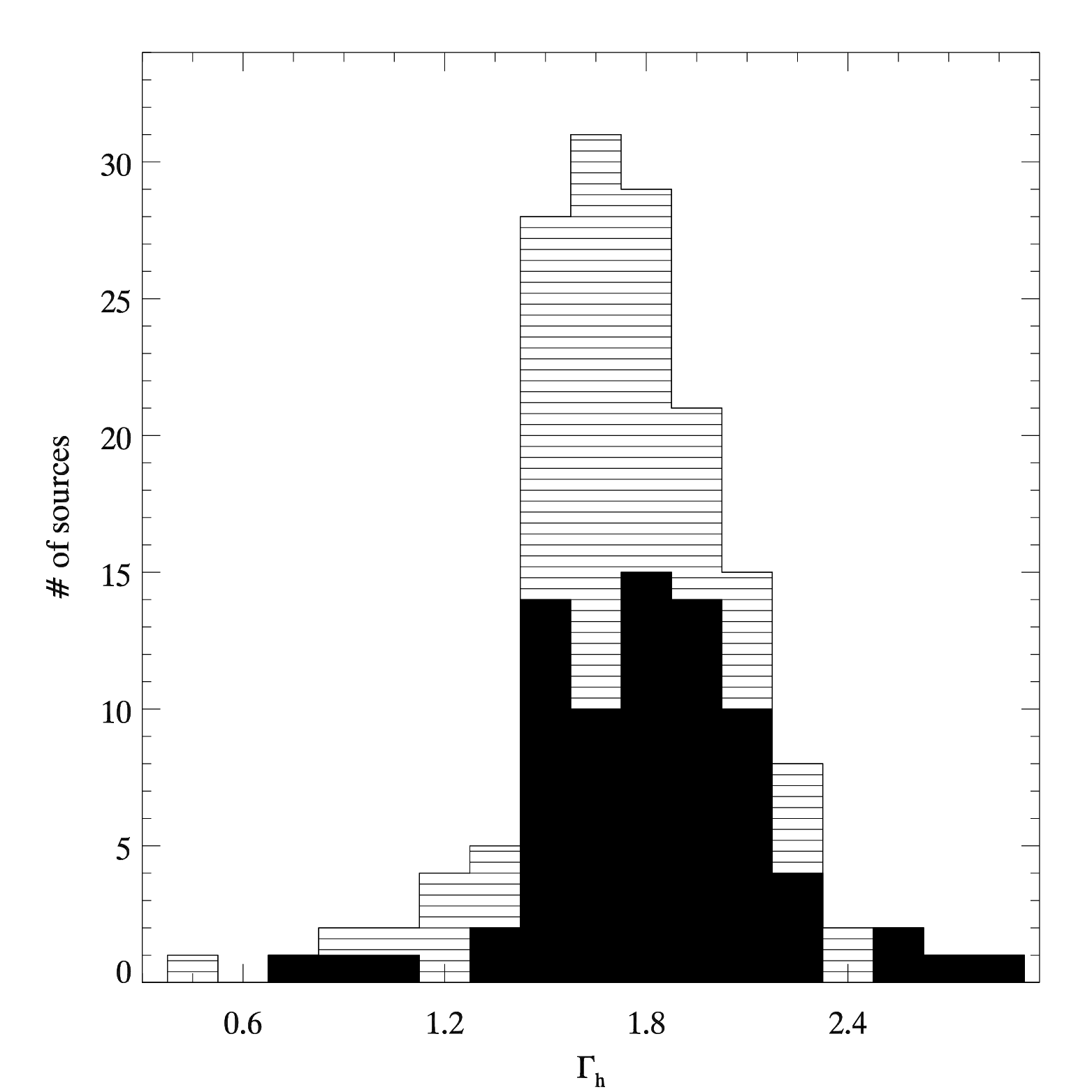, width=8cm, height=8cm}
\epsfig{file=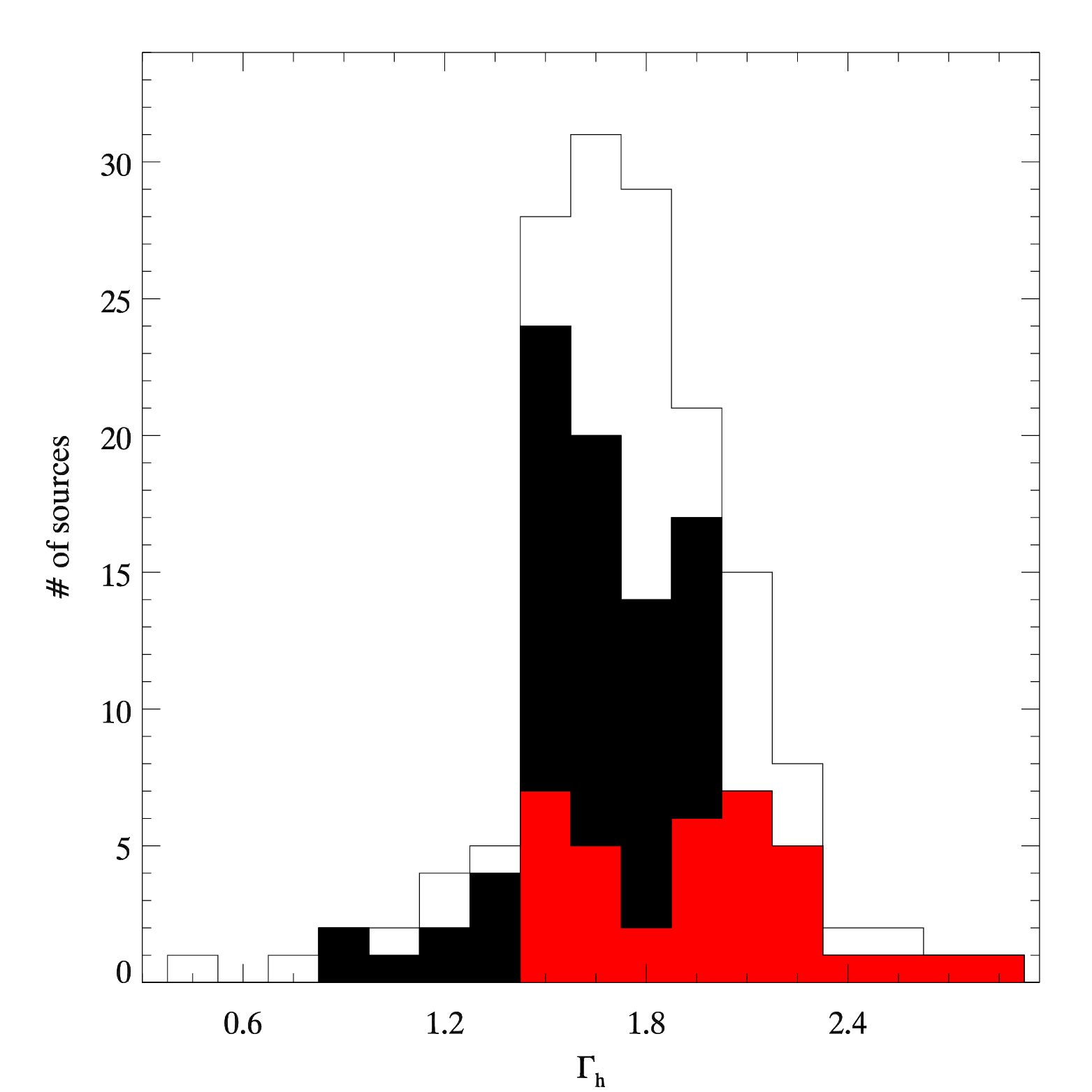, width=8cm, height=8cm}
\epsfig{file=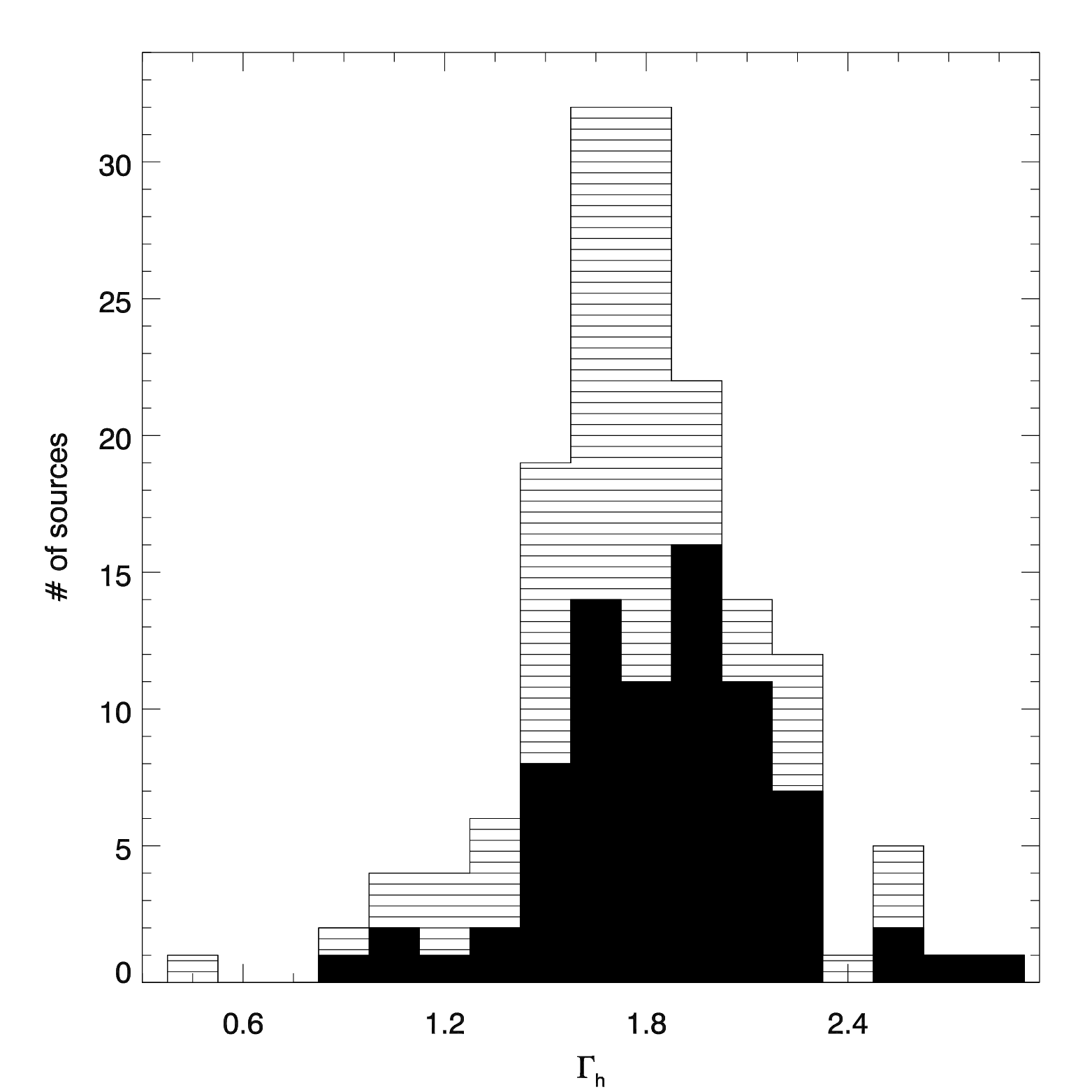, width=8cm, height=8cm}
\epsfig{file=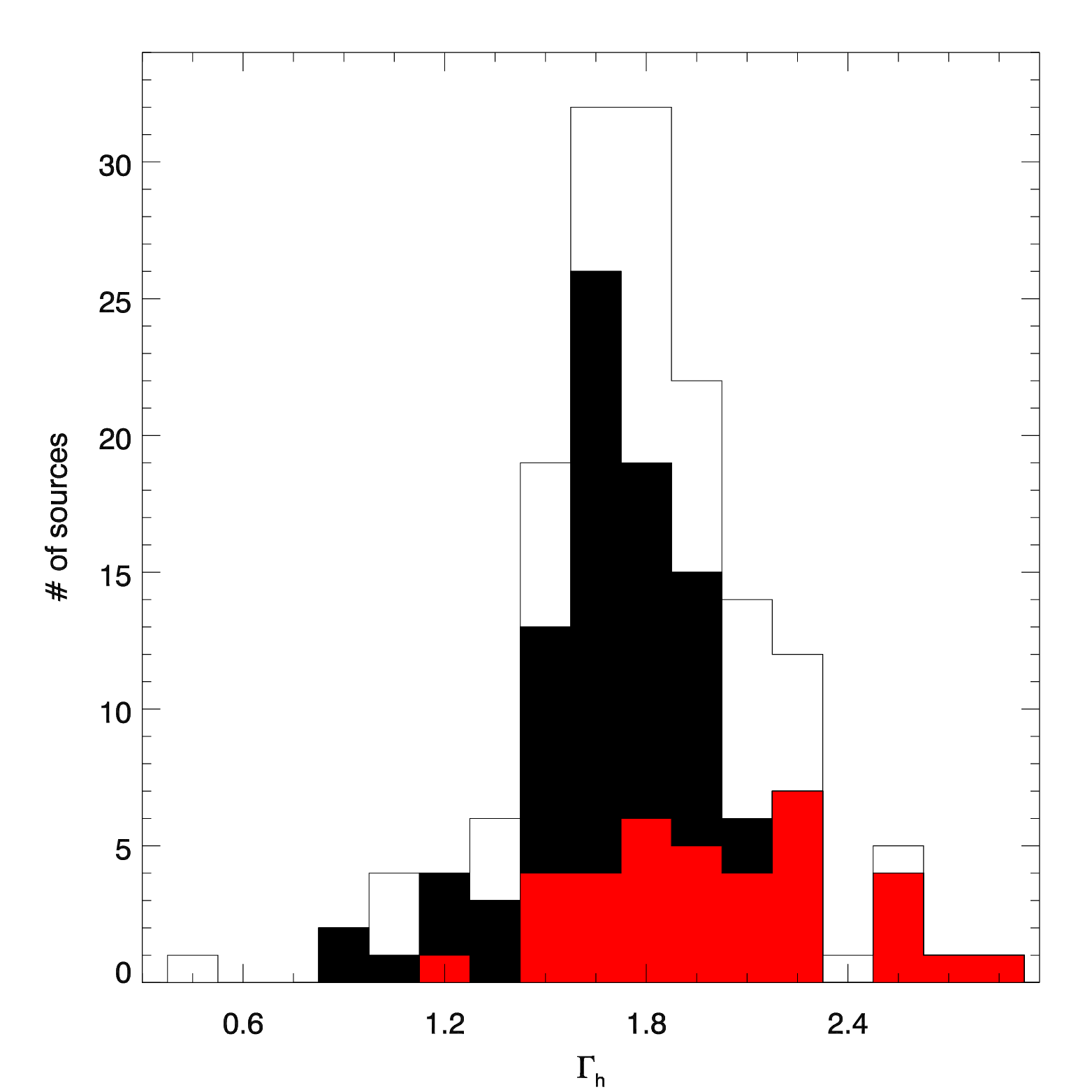, width=8cm, height=8cm}
\end{center}
\caption{\label{gammadist}Hard X-ray powerlaw index distribution in CAIXA, for the best fit model (\textit{upper panels}) and the model which includes a luminosity-dependent Compton reflection component (\textit{lower panels}). On the left, dashed histograms are Seyfert galaxies, filled ones are quasars. On the right, red histograms are narrow-line objects, black histograms are broad-line objects, while white histograms are used for sources without a measure of the H$\beta$ FWHM in CAIXA. See text for details.}
\end{figure*}

The average\footnote{These mean values are `censored', because they take into account errors and upper limits on the measures, using the method described in Appendix \ref{appa}.} 2-10 keV spectral index for the whole catalogue is $1.73\pm0.04$, with a large spread of $\sigma=0.45\pm0.03$. The hard X-ray $\Gamma$ appears steeper in quasars \citep[$1.80\pm0.05$, consistent with the value obtained for the PG quasars:][]{pico05} than in Seyferts ($1.66\pm0.05$). A Kolmogorov-Smirnov test yields that the two populations are different at the 99\% confidence level. The difference is even larger if we consider narrow-line objects ($\Gamma=1.94\pm0.07$) with respect to broad-line objects ($1.62\pm0.04$). In this case, a K-S test gives a NHP of $2\times10^{-5}$: the two populations are significantly different with respect to their X-ray photon indexes. It is important to stress here that the photon indexes quoted in CAIXA, in principle, are not the intrinsic ones, because in our fits we do not consider the effect of Compton reflection and warm absorption. Therefore, the reader should keep in mind that the intrinsic photon indexes are likely to be steeper than the ones reported here, as in the case presented in the previous section. Indeed, when the Compton reflection component can be well constrained, as in broad band \textit{BeppoSAX} spectra, the average spectral index is steeper \citep[$1.89\pm0.03$ for Seyfert galaxies: ][]{dad08}. It is, therefore, possible that the different photon indexes found for different populations in CAIXA do not only (or at all) reflect differences in the intrinsic $\Gamma$ of the sources, but also on the amount of Compton reflection.

In order to test this hypothesis, we performed a second run of automating fitting procedure, with a model which includes Compton reflection. Since the amount of reflection cannot be constrained with the EPIC pn band coverage (and this is the reason why it is not included in the default model adopted in CAIXA) we fixed it in each source to a value tightly linked with the EW of the neutral iron K$\alpha$ line. A linear correlation between the two parameters was assumed, adopting R=$\frac{\Omega}{2\pi}=1$ for EW=120 eV as normalization \citep[see e.g.][]{mpp91}. Furthermore, in order to avoid the technical issues involved in a fit with R linked to the EW of an emission line, we derived the expected iron K$\alpha$ EW on the basis of the hard X-ray luminosity, following the best fit for the Iwasawa-Taniguchi effect \citep{bianchi07}:

\begin{equation}
\label{iteffect}
\log(EW_{Fe}) = \left(1.73\pm0.03\right) + \left(-0.17\pm0.03\right) \log(L_{X,44})
\end{equation}

\noindent where $EW_{Fe}$ is the EW of the neutral iron K$\alpha$ line in eV and $L_{X,44}$ is the 2-10 keV X-ray luminosity in units of $10^{44}$ erg s$^{-1}$. We, therefore, introduce a luminosity-dependent Compton reflection component. As for the default model in CAIXA, we added a second powerlaw to model the soft X-ray emission, only if statistically required.

The distributions of the hard X-ray powerlaw index derived for the model which includes Compton reflection are shown in the bottom panels of Fig. \ref{gammadist}. The average 2-10 keV spectral index for the whole catalogue is, as expected, steeper, but not significantly, being now $1.78\pm0.04$ ($\sigma=0.46\pm0.03$). Quasars ($1.83\pm0.05$) and Seyferts ($1.73\pm0.05$) are still different at the 99\% level, on the basis of a K-S test. Narrow-line ($\Gamma=2.00\pm0.07$) and broad-line objects ($1.65\pm0.04$) still display a very significant difference (NHP of $4\times10^{-5}$ for the K-S test). The inclusion of a luminosity-dependent Compton reflection component, therefore, does not significantly affect the distribution of the powerlaw indexes in CAIXA and, more importantly, does not alter the differences between populations found with the simpler model.

Another important component which is not included in the default model adopted in CAIXA is the warm absorber, which is known to be present in many sources \citep[see e.g.][]{pico04}. If it is not properly modeled, it may result in a flatter powerlaw index, as in the case of NGC~3783 reported in Sect. \ref{goodfit}. However, a correct modelisation of the complex (and often multiple) warm absorption component is far beyond the scopes of CAIXA and will not be addressed any more in the following. Nevertheless, we would like to stress that most of the studies on large samples of AGN adopt simple models, so that the hard X-ray powerlaw index reported in CAIXA can be directly compared to the ones found in literature. On the other hand, it is clear that the value of the intrinsic $\Gamma$ for single sources may significantly differ from those found with more complex models, but this is inherent in any statistical analysis of catalogues and the effect is diluted by the large numbers of sources.

\subsection{\label{iron}The iron lines}

\begin{figure*}
\begin{center}
\epsfig{file=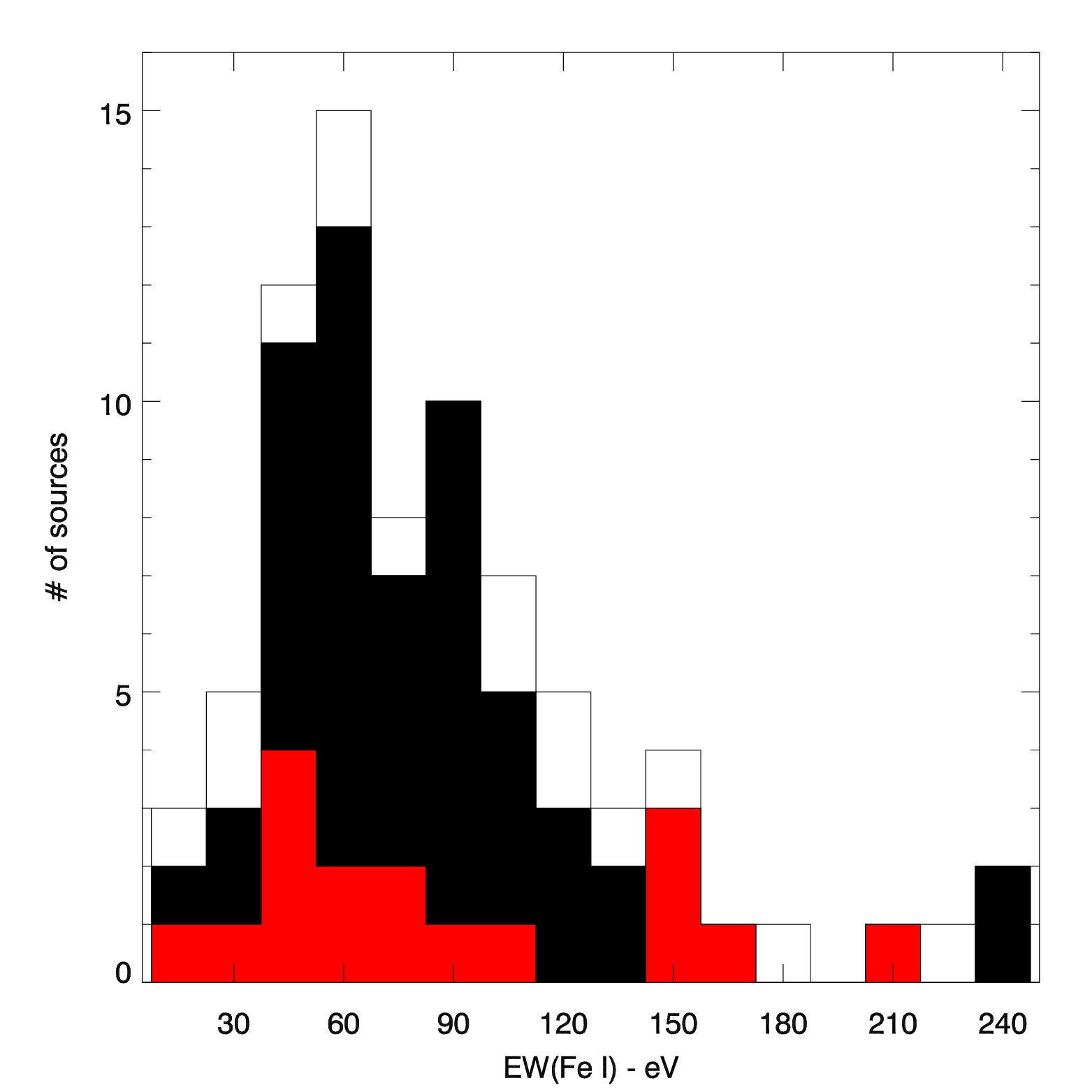, width=6cm}
\epsfig{file=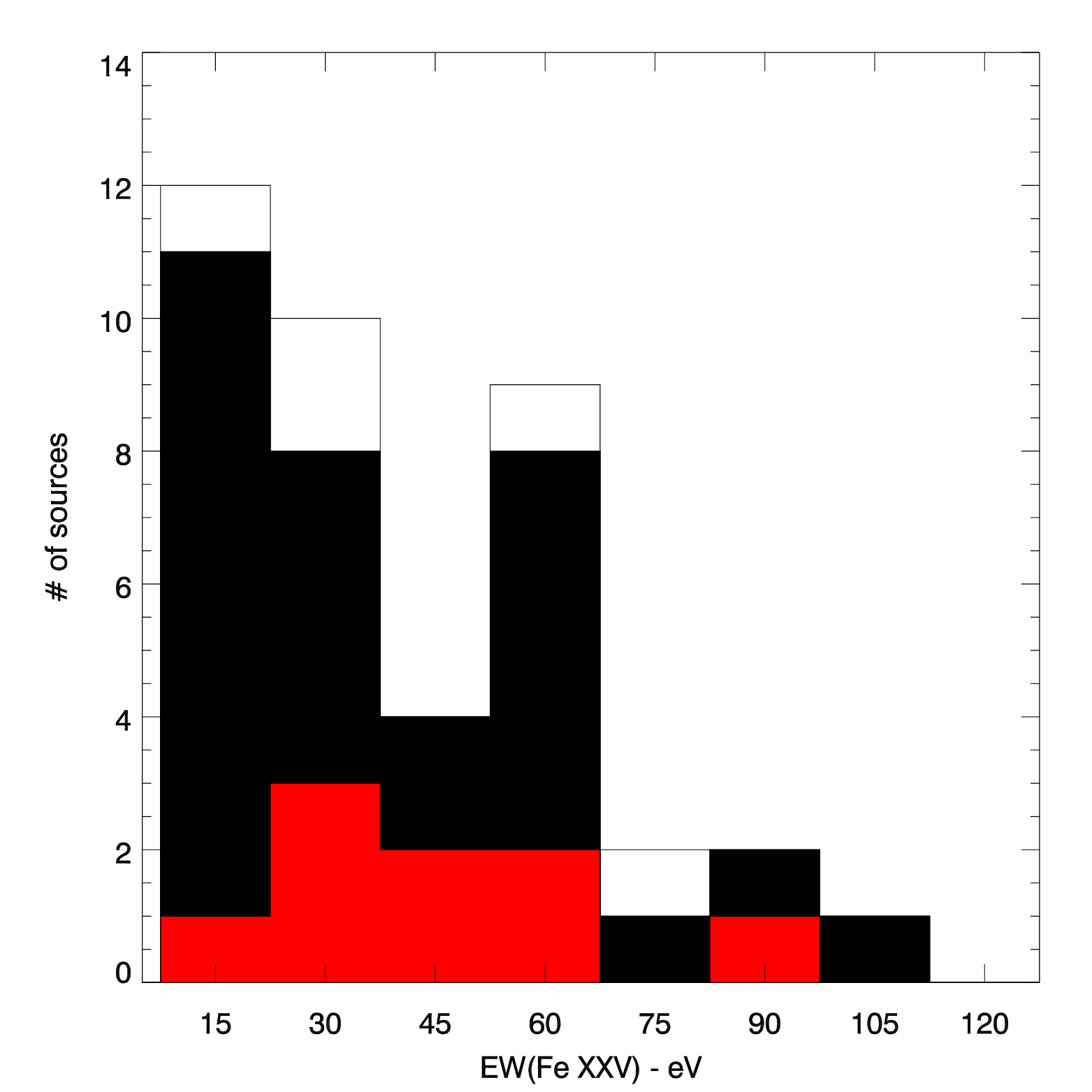, width=6cm}
\epsfig{file=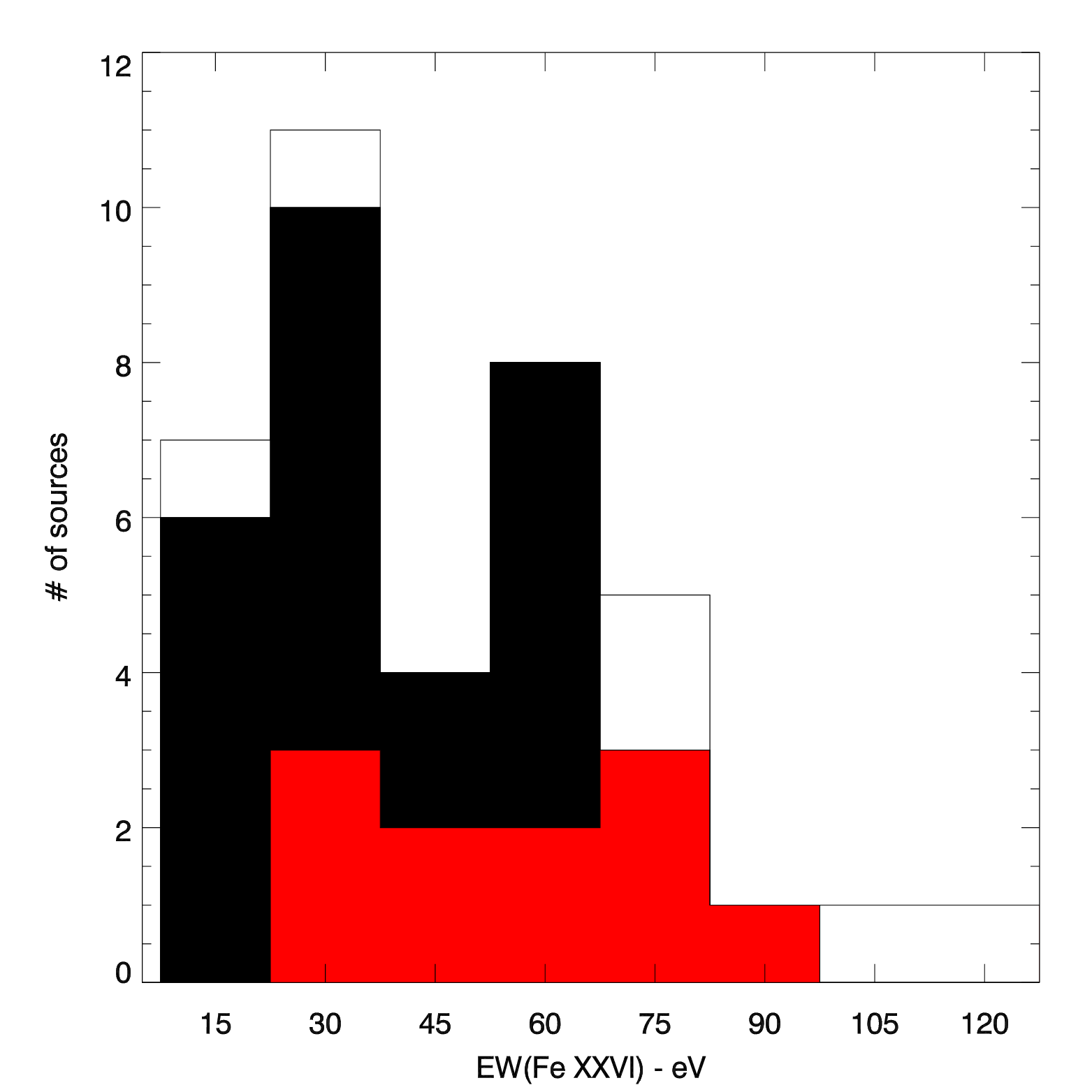, width=6cm}
\end{center}
\caption{\label{fe2526}Distribution of neutral iron, Fe \textsc{xxv} (\textit{left}) and Fe \textsc{xxvi} (\textit{right}) measured (no upper limits) EWs in CAIXA. Red histograms are narrow-line objects, black histograms are broad line objects and white histograms are for objects with no measure in CAIXA for the H$\beta$ FWHM. Very large EWs are not included in these distributions. See text for details.}
\end{figure*}

The model adopted to fit the X-ray spectra in CAIXA includes the K$\alpha$ emission lines from neutral iron, Fe \textsc{xxv} and Fe \textsc{xxvi} (see Sect. \ref{data}). In Fig. \ref{fe2526}, we show the distributions of the EW for the three iron lines. The neutral iron line is detected in 81 sources, 46 of which are broad-line objects (out of the 65 in CAIXA) and 18 narrow-line objects (out of the 37 in CAIXA)\footnote{These values are slightly different from the ones reported in \citet{bianchi07}, being 81, 43 and 19, because of the update of the H$\beta$ FWHM values in CAIXA.}. The distribution is clustered around 50-120 eV, as expected if arising as reprocessing from cold Compton-thick matter, as the putative torus \citep[see e.g.][]{mpp91}. Only 4 sources have (detected) EWs larger than 250 eV, while the average (including upper limits) value of the EW is $76\pm6$ eV, with quite a large spread ($\sigma=75$ eV). Such a large spread is also expected, since the neutral iron K$\alpha$ EW is strongly anti-correlated to the X-ray luminosity \citep[the Iwasawa-Taniguchi effect:][]{bianchi07}.

The Fe \textsc{xxv} line is detected in 44 sources, 26 of which are broad-line objects and 11 narrow-line objects (out of the 37 in CAIXA). Similarly, the Fe \textsc{xxvi} line is detected in 40 sources, 21 broad-line objects and 12 narrow-line objects. We therefore do not find different rates of detection of ionised iron lines for these two classes of AGN. These lines may be produced in a Compton-thin gas photoionised by the AGN, but EWs larger than $\simeq100$ eV are quite difficult to explain within this scenario \citep[see e.g.][]{bm02,bianchi05}. The effect of the neutral K$\beta$ line which may, in principle, contaminate the flux of the Fe \textsc{xxvi} line, is properly taken into account in CAIXA. However, note that lines with such a large value of EW are few in CAIXA and always with very large errors (see Fig. \ref{fe2526}, considering that EWs larger than 130 eV are detected only in 4 and 2 sources, for Fe \textsc{xxv} and Fe \textsc{xxvi}, respectively). In any case, we cannot exclude that an ionised accretion disc may contribute to the flux of the ionised iron lines in the objects where the measured EWs are quite large, as suggested, for example, in Mrk~766 by variability studies \citep{mill06}.

\begin{figure}
\begin{center}
\epsfig{file=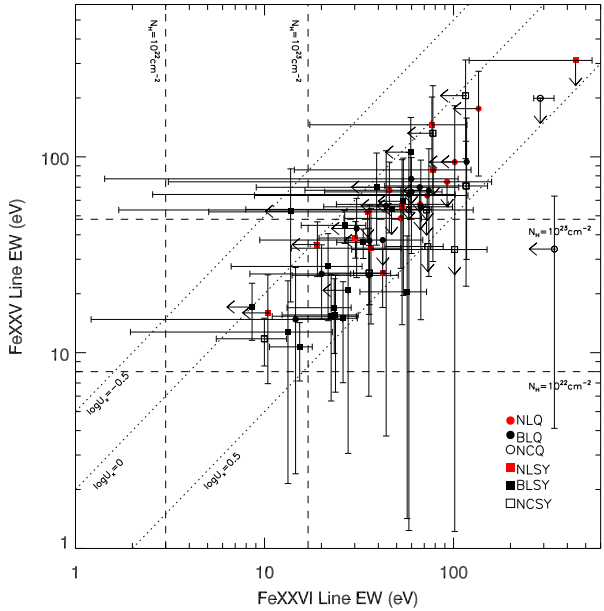, width=8.5cm}
\end{center}
\caption{\label{ionizediron}Fe \textsc{xxv} against Fe \textsc{xxvi} EWs for the objects in CAIXA where at least one of the emission lines is detected (upper limits for both lines are excluded). The diagonal lines shows the expected ratios between the two for three different values of the X-ray ionization parameter \citep[see][for details]{bm02}. The dashed lines show the EWs expected for the two lines if they are produced in a gas photoionised by the AGN \citep[see][for details]{bm02}. The different symbols refers to the classification of the objects, on the basis of their absolute magnitude and H$\beta$ FWHM: \textit{NLSY}, narrow-line Seyfert 1; \textit{BLSY}, broad-line Seyfert 1; \textit{NCSY}, not-classified Seyfert 1 (no H$\beta$ FWHM measure available); \textit{NLQ}, narrow-line quasar; \textit{BLQ}, broad-line quasar; \textit{NCQ}, not-classified quasar (no H$\beta$ FWHM measure available).}
\end{figure}

Both the Fe \textsc{xxv} and Fe \textsc{xxvi} lines are detected in 24 sources, 15 broad-line objects and 6 narrow-line objects. We plot in Fig. \ref{ionizediron} the EW of one line against the other. The ratio between the two EWs is in agreement with photoionised models, but on average pointing towards large values of the ionization parameter \citep{bm02}. This is not surprising, given the fact that these sources are those where both lines are observed and thus the gas must be ionised enough to produce a large EW for H-like iron. Column densities up to $\simeq10^{23}$ cm$^{-2}$ are required to reproduce most of the observed EWs for He-like iron, while the column densities suggested by the Fe \textsc{xxvi} EWs are, on average, larger. This result is not uncommon: see \citet{bianchi05} for a detailed discussion. 

\subsection{The soft excess}

As mentioned in Sect. \ref{goodfit}, the presence of a second powerlaw to model the soft X-ray emission is statistically required in 123 spectra out of 156. Although the need for this component is clearly requested in the presence of neutral absorption of the primary continuum (which is present only in 14 sources), in the majority of the cases it reflects the presence of an excess in the soft X-ray regime with respect to the intrinsic powerlaw emission. Considering also that most of the single-powerlaw fits are clustered in the low-SNR spectra, where the presence of a soft excess is more difficult to detect, we can conclude that this component is very common in  CAIXA, in agreement with previous studies \citep[see e.g.][]{pico04}.

\begin{figure*}
\begin{center}
\epsfig{file=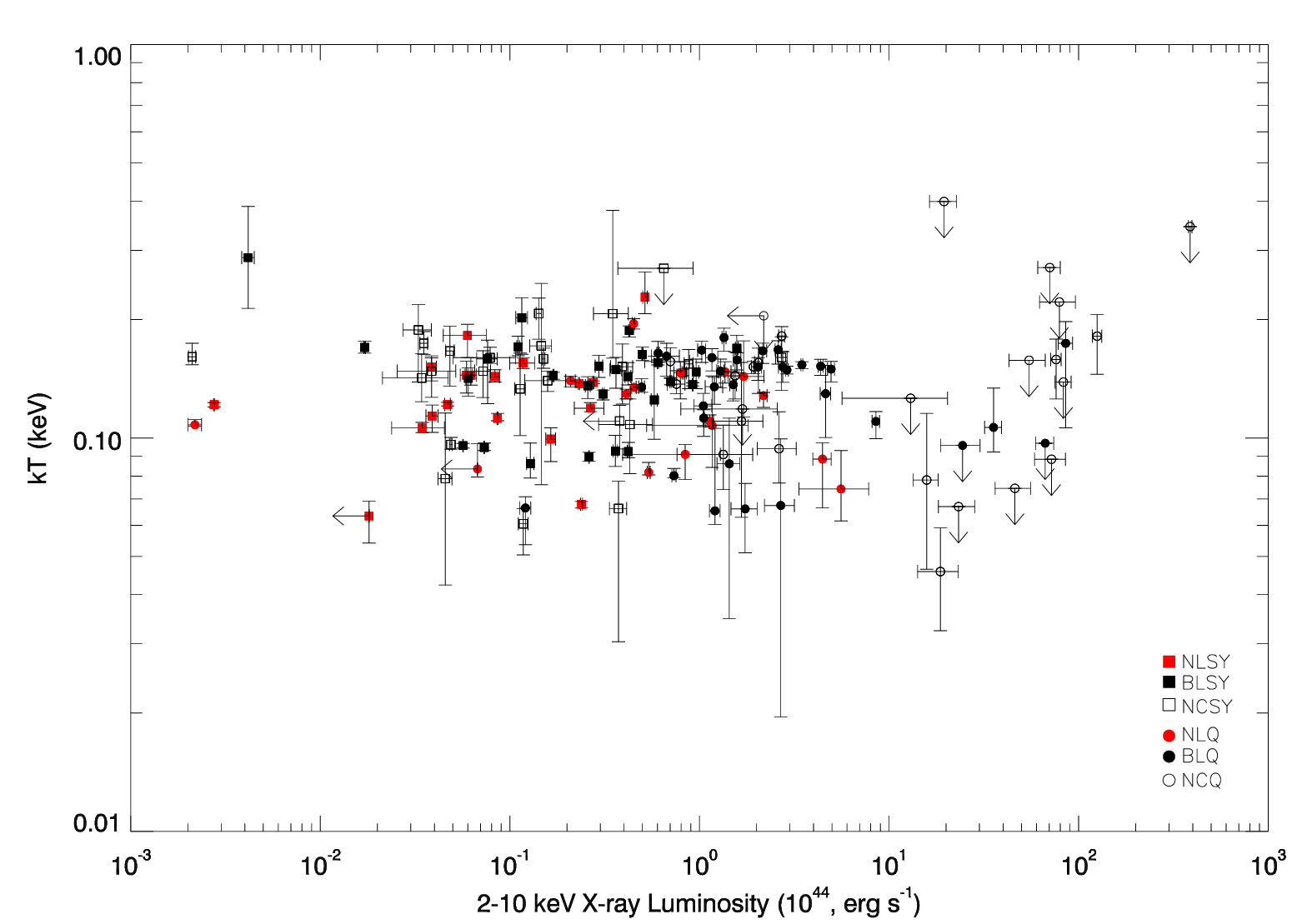, width=8.5cm}
\hspace{0.5cm}
\epsfig{file=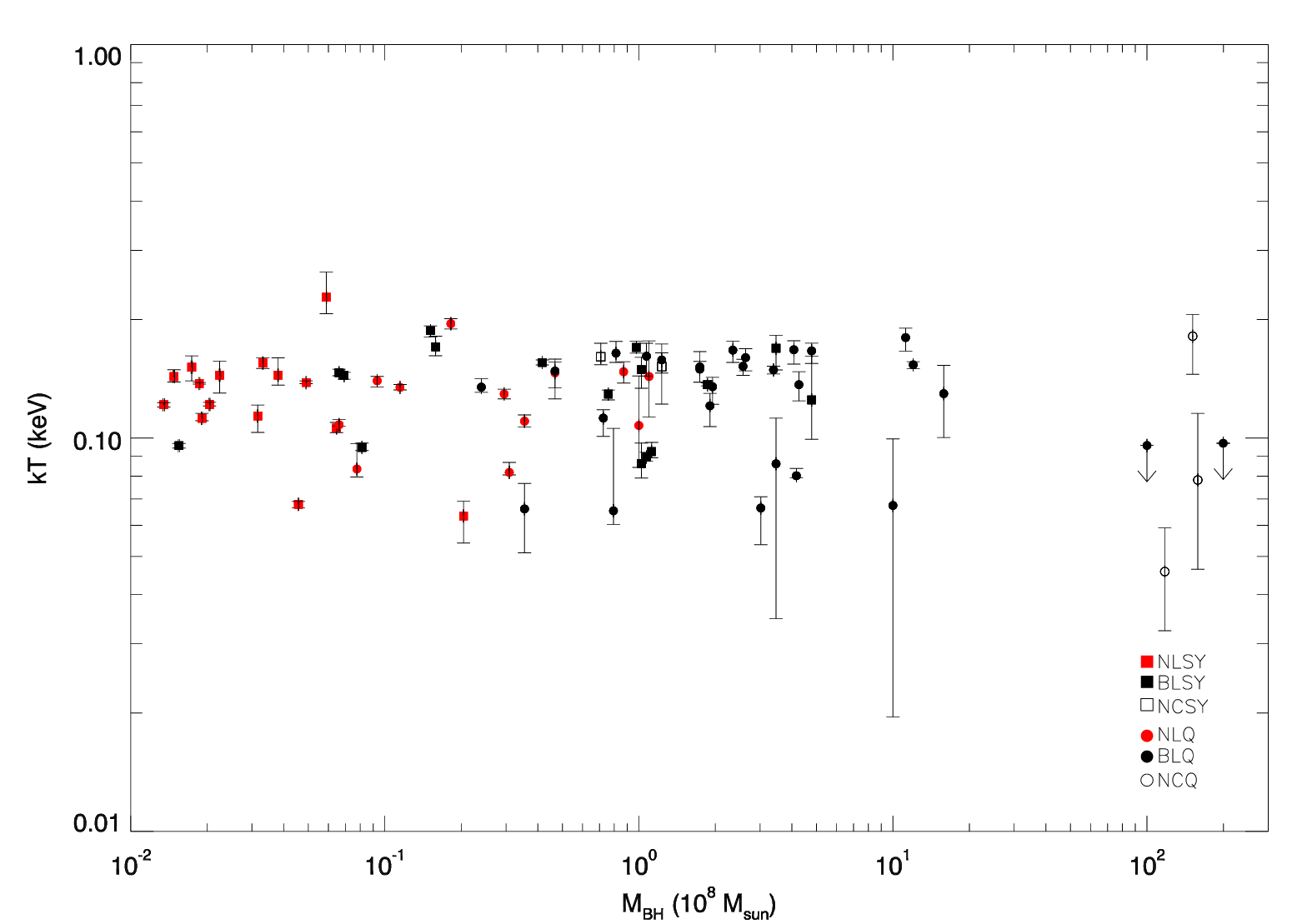, width=8.5cm}
\end{center}
\caption{\label{bbkt}Temperature of the black body model, when used to model the soft excess in the objects of CAIXA. It appears completely unrelated to the X-ray luminosity (\textit{left} and the BH mass (\textit{right}). The different symbols refers to the classification of the objects, on the basis of their absolute magnitude and H$\beta$ FWHM: \textit{NLSY}, narrow-line Seyfert 1; \textit{BLSY}, broad-line Seyfert 1; \textit{NCSY}, not-classified Seyfert 1 (no H$\beta$ FWHM measure available); \textit{NLQ}, narrow-line quasar; \textit{BLQ}, broad-line quasar; \textit{NCQ}, not-classified quasar (no H$\beta$ FWHM measure available).}
\end{figure*}

The modelisation of the soft excess with a simple powerlaw has no physical justification. Indeed, the values of the soft X-ray indexes found in CAIXA are often very steep (see Table \ref{xraydata2}) and cannot be naturally explained. Moreover, as already showed by several authors \citep{gd04,crummy06}, the soft excess in unobscured radio-quiet AGN, when fitted with a thermal model, is characterized by a temperature which is amazingly constant across orders of magnitude of luminosities and BH masses. We tested the same scenario in CAIXA, adopting another fitting model, with a thermal black body component instead of the soft powerlaw. The results are shown in Fig. \ref{bbkt}: even on a catalogue of more sources and which spans larger orders of magnitude in luminosities and BH masses than previous studies, the temperature of the thermal emission appears completely unrelated to the fundamental properties of the AGN. Moreover, it was shown that, even for the same object, a variation of the soft X-ray luminosity of one order of magnitude does not produce any variation of this `universal' temperature, in contrast with what expected from the basic physics of disk accretion and black bodies \citep{ponti06}.

In the last few years, two main competing models have been proposed to explain the origin of the soft excess, both ascribing the apparent universal temperature of the soft excess to atomic physics processes. In one scenario, the soft excess arises from the enhancement of reflection from the inner regions of the accretion disk due to light-bending effects, together with a strong suppression of the primary emission \citep[e.g.][]{mf04}. On the other hand, the soft excess can also be mimicked by absorption from a relativistically outflowing warm gas \citep[e.g.][]{gd04}. Present X-ray spectra do not allow us to discriminate between the two models.

Therefore, we decided to investigate the nature of the soft excess adopting a model-independent parameter. We chose the ratio between the soft X-ray luminosity (rest-frame 0.5-2 keV) and the hard X-ray luminosity (rest-frame 2-10 keV) as an indicator of the importance of the soft excess with respect to the overall X-ray spectrum of the source. As shown in Sect. \ref{goodfit}, the luminosities extracted with the automatic fitting procedure in CAIXA are not significantly affected by the adopted model and so their ratio is a good parameter to characterize at least the basic property of the soft excess of the sources, i.e. its strength with respect to the primary X-ray emission.

\begin{figure}
\begin{center}
\epsfig{file=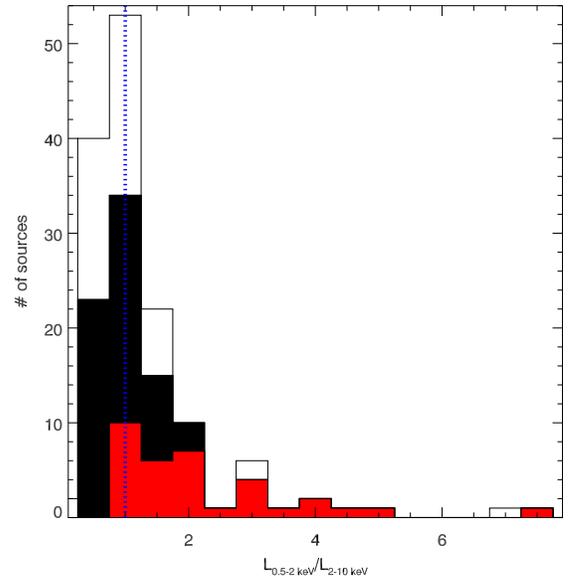, width=8cm, height=8cm}
\end{center}
\caption{\label{xratiodist}Hard X-ray powerlaw index distribution in CAIXA. Red histograms are narrow-line objects, black histograms are broad line objects and white histograms are for objects with no measure in CAIXA for the H$\beta$ FWHM. See text for details.}
\end{figure}

In Fig. \ref{xratiodist} we show the distribution of the X-ray luminosity ratio in CAIXA. As already mentioned in Sect. \ref{data}, we stress here that this value is calculated only for sources with no neutral absorption, for which the intrinsic soft X-ray luminosity is not measurable. On the other hand, as already commented in previous sections, a warm absorber component is not modeled in CAIXA, so some of the calculated X-ray luminosity ratios may be lower than the intrinsic ones: in some objects, low ratios may not be an intrinsic property, but instead they may indicate the presence of a strong warm absorption component. Although the majority of the sources cluster around a value of 1, narrow- and broad-line objects are significantly different populations with respect to the X-ray luminosity ratio, the NHP being $6\times10^{-9}$ for a K-S test. Indeed, it is very interesting to note that no narrow-line object is found with a value of the luminosity ratio lower than unity. In other words, narrow-line objects tend to have a strong soft X-ray luminosity with respect to their hard X-ray emission. We defer the reader to a forthcoming companion CAIXA paper for a detailed discussion on the correlations between the X-ray luminosity ratio and other multiwavelength properties of AGN, and its possible consequences on the origin of the soft excess.

\section{Conclusions}

We presented CAIXA, a Catalogue of AGN In the XMM-\textit{Newton} Archive. It consists of all the radio-quiet X-ray unobscured ($\mathrm{N_H}<2\times10^{22}$ cm$^{-2}$) AGN observed by XMM-\textit{Newton} in targeted observations, whose data are public as of March 2007. We performed a complete and homogeneous spectral analysis of the X-ray data, whose main results can be summarised as follows:

\begin{itemize}

\item \textit{The X-ray spectral index.} The average 2-10 keV spectral index for the whole catalogue is $1.73\pm0.04$ ($\sigma=0.49\pm0.02$) and appears significantly steeper in narrow-line objects and quasars with respect to broad-line objects and Seyfert galaxies. The average value for CAIXA and the differences between source populations do not significantly change if a luminosity-dependent Compton reflection component is included in the model.

\item \textit{The iron lines.} The distribution of the neutral iron K$\alpha$ EW is clustered around 50-120 eV, as expected if arising as reprocessing from cold Compton-thick matter, as the putative torus. Fe \textsc{xxv} and Fe \textsc{xxvi} are commonly detected, but we do not find different rates of detection of these lines between narrow- and broad-line objects. The ratio between the EWs of the two lines is, on average, broadly in agreement with photoionisation models.

\item \textit{The soft excess.} This component is very common in  CAIXA, in agreement with previous studies. We chose the ratio between the soft X-ray luminosity (rest-frame 0.5-2 keV) and the hard X-ray luminosity (rest-frame 2-10 keV) as a model-independent indicator of the strength of the soft excess. With respect to this parameter, narrow- and broad-line objects are significantly different populations. Indeed, it is very interesting to note that no narrow-line object is found with a value of the luminosity ratio lower than unity.

\end{itemize}

On the basis of these results, we will investigate the correlations between the X-ray and the multiwavelength properties of the sources in CAIXA in a companion paper.

\acknowledgement
SB and GM acknowledge financial support from ASI (grant I/088/06/0), GP from ANR (ANR-06-JCJC-0047). We acknowledge support from the Faculty of the European Space Astronomy Centre (ESAC). We thank C. Gordon for all his efforts in solving problems related to \textsc{Xspec} and the anonymous referee for his/her suggestions. Based on observations obtained with XMM-Newton, an ESA science mission with instruments and contributions directly funded by ESA Member States and NASA.

\begin{appendix}

\section{\label{appa}Censored mean}

Given the large errors and the presence of upper limits for many of the parameters derived in CAIXA, most average values cited in this paper are `censored means', if not otherwise stated. The adopted method is is very similar to the `censored fits' presented in \citet{bianchi07}, based on the linear fits performed by \citet{gua06}, and can be summarized as follows. A large number of mean values was performed on a set of Monte-Carlo simulated data derived from the experimental points according to the following rules: a) each detection was substituted by a random Gaussian distribution, whose mean is the best-fit measurement and whose standard deviation is its statistical uncertainty; b) each upper limit $U$ was substituted by a random uniform distribution in the interval [0,U]. The average (with relative uncertainty) of the mean values of each single data set is the `censored mean' reported in the text, along with the average dispersion. 

\end{appendix}

\bibliographystyle{aa}
\bibliography{sbs}

\longtab{1}{
\begin{longtable}{ccccccc}
\caption{\label{xraydata}Main properties of the observations of the sources included in CAIXA.}\\
\hline \hline
\textbf{Source} & \textbf{ObsID} & \textbf{T (ks)} &\textbf{F(0.5-2 keV)} & \textbf{F(2-10 keV)} &   \textbf{L(0.5-2 keV)}&  \textbf{L(2-10 keV)} \\
(1) & (2) & (3) & (4) & (5) & (6) & (7)\\
\hline
 &  & & & & &\\
\endfirsthead
\caption{continued.}\\
\hline \hline
\textbf{Source} & \textbf{ObsID} & \textbf{T (ks)} &\textbf{F(0.5-2 keV)} & \textbf{F(2-10 keV)} &   \textbf{L(0.5-2 keV)}&  \textbf{L(2-10 keV)} \\
(1) & (2) & (3) & (4) & (5) & (6) & (7)\\
\hline
 &  & & & & &\\
\endhead
\hline
\endfoot
..&..  &.. &.. &.. &.. &..\\
&  & & & & &\\
\hline

\end{longtable}
\textit{Full table will be published in electronic form at the CDS.}

(1) Source name \citep{ver06}; (2) XMM-\textit{Newton} \textsc{obsid} (3) Clean exposure time;  (4) 0.5-2 keV flux ($10^{-12}$ erg cm$^{-2}$ s$^{-1}$); (5) 2-10 keV flux ($10^{-12}$ erg cm$^{-2}$ s$^{-1}$); (6) Unabsorbed 0.5-2 keV luminosity ($10^{44}$ erg s$^{-1}$); (7): Unabsorbed 2-10 keV luminosity ($10^{44}$ erg s$^{-1}$).
}

\longtab{2}{
\begin{longtable}{ccccccc}
\caption{\label{xraydata2}Main X-ray parameters derived from the automatic fits of the observations included in CAIXA.}\\
\hline \hline
\textbf{Source} & \textbf{Obsid} & $\mathbf{\Gamma_s}$  & $\mathbf{\Gamma_h}$ & $\mathbf{EW_{6.4}}$  & $\mathbf{EW_{6.7}}$ & $\mathbf{EW_{6.96}}$ \\
(1) & (2) & (3) & (4) & (5) & (6) & (7)\\
\hline
& &  & & & & \\
\endfirsthead
\caption{continued.}\\
\hline \hline
\textbf{Source} & \textbf{Obsid} & $\mathbf{\Gamma_s}$ & $\mathbf{\Gamma_h}$ & $\mathbf{EW_{6.4}}$  & $\mathbf{EW_{6.7}}$ & $\mathbf{EW_{6.96}}$ \\
(1) & (2) & (3) & (4) & (5) & (6)& (7) \\
\hline
 &  & & & & &\\
\endhead
\hline
\endfoot
..&.. &..  &.. &.. &.. &.. \\
& &  & & & & \\
\hline

\end{longtable}
\textit{Full table will be published in electronic form at the CDS.}

(1) Source name \citep{ver06}; (2) XMM-\textit{Newton} \textsc{obsid}; (3) Soft X-ray powerlaw index; (4) Hard X-ray powerlaw index; (5) 6.4 keV emission line EW (eV); (6) 6.7 keV emission line EW (eV); (7) 6.96 keV emission line EW (eV).
}

\longtab{3}{
\begin{longtable}{ccccccccccc}
\caption{\label{multitable}Multiwavelength properties of the sources included in CAIXA. RQ means the source is consided radio-quiet in the reference, but no radio flux is reported.}\\
\hline \hline
\textbf{NAME} & \textbf{RA} & \textbf{DEC} &  \textbf{z} & $\mathbf{M_{abs}}$ & $\mathbf{F_{6cm}}$ & $\mathbf{F_{20cm}}$ & $\mathbf{H_{\beta}}$ & $\mathbf{M_{BH}}$ & $\mathbf{m_{v}}$ & \textbf{B-V}\\
(1) & (2) & (3) & (4) & (5) & (6) & (7) & (8) & (9) & (10) & (11)\\
\hline
 &  & & & & &&&&&\\
\endfirsthead
\caption{continued.}\\
\hline \hline
\textbf{NAME} & \textbf{RA} & \textbf{DEC} &  \textbf{z} & $\mathbf{M_{abs}}$ & $\mathbf{F_{6cm}}$ & $\mathbf{F_{20cm}}$ & $\mathbf{H_{\beta}}$ & $\mathbf{M_{BH}}$ & $\mathbf{m_{v}}$ & \textbf{B-V}\\
(1) & (2) & (3) & (4) & (5) & (6) & (7) & (8) & (9) & (10) & (11)\\
\hline

.. & .. & .. & .. & .. & .. & .. & .. & .. & .. &\\

\endhead
\hline
\endfoot

.. & .. & .. & .. & .. & .. & .. & .. & .. & .. &\\
 &  &  &  &  &  &  &  &  &  &\\
\hline

\end{longtable}
\textit{Full table will be published in electronic form at the CDS.}

(1) Source name \citep{ver06}; (2)-(3): Coordinates in J2000 (NED);  (4) Redshift \citep{ver06}; (5) Absolute magnitude \citep{ver06}; (6)-(7): Radio fluxes at 6 and 20 cm (Jy). References: 1. \citet{ver06}, 2. \citet{kel89}, 3. \citet{greg96}, 4. \citet{nraosur}, 4u. Sensitivity limit in \citet{nraosur}, 5. \citet{kuhn01}, 6. \citet{veron91}, 7. \citet{gallo04}, 8. \citet{sw80}, 9. \citet{page04}, 10. \citet{gan06}, 11. \citet{mauch03}, 12. \citet{yuan98}, 13. \citet{wang04} 14. \citet{sad02}, 15. , 16. \citet{mcm99}; (8) FWHM (km s$^{-1}$). References: a. \citet{bg92}, b. \citet{wl01}, c. \citet{vest02}, d. \citet{wpm99}, e. \citet{bbf96}, f. \citet{grupe99} g. \citet{grupe04}, h. \citet{bz03}, i. \citet{pet04}, j. \citet{wang96}, k. \citet{leig99}, l. \citet{marz03}, m. \citet{mcsh99}, n. \citet{win92}, o. \citet{gan01}, p. \citet{corb91}, q. \citet{cb96}, r. \citet{shem06}, s. \citet{ckg03}, t. \citet{md01}, u. \citet{laor00}, v. \citet{rey97b}, w. \citet{zw05}; (9) Log of BH mass in units of solar masses. References: A. \citet{wu02}, B. \citet{vp06}, C. \citet{vest02}, D. \citet{zw06}, E. \citet{wl01}, F. \citet{shem06}, G. \citet{cze04}, H. \citet{laor98}, I. \citet{bian03}, J. \citet{oneill05}, K. \citet{merl03}, L. \citet{zw05} - Notes $r$:reverberation mapping, $o$: H$\beta$ FWHM, $m$: MgII FWHM, $v$: stellar velocity dispersion; $n$: narrow line region size (10) $m_V$ \citep{ver06}; (11) B-V (mag) \citep{ver06}.
}

\end{document}